# Communication and Personality Profiles of Global Software Developers

Sherlock A. Licorish and Stephen G. MacDonell
Department of Information Science
University of Otago
PO Box 56, Dunedin 9054
New Zealand
sherlock.licorish@otago.ac.nz, stephen.macdonell@otago.ac.nz

## ABSTRACT

**Context**: Prior research has established that a small proportion of individuals dominate team communication during global software development. It is not known, however, how these members' contributions affect their teams' knowledge diffusion process, or whether their personality profiles are responsible for their dominant presence. **Objective**: We set out to address this gap through the study of repository artifacts. **Method**: Artifacts from ten teams were mined from the IBM Rational Jazz repository. We employed social network analysis (SNA) to group practitioners into two clusters, *Top Members* and *Others*, based on the numbers of messages they communicated and their engagement in task changes. SNA metrics (density, in-degree and closeness) were then used to study practitioners' importance in knowledge diffusion. Thereafter, we performed psycholinguistic analysis on practitioners' messages using linguistic dimensions that had been previously correlated with the Big Five personality profiles. **Results**: For our sample of 146 practitioners we found that Top Members occupied critical roles in knowledge diffusion, and demonstrated more openness to experience than the Others. Additionally, all personality profiles were represented during teamwork, although openness to experience, agreeableness and extroversion were particularly evident. However, no specific personality predicted members' involvement in knowledge diffusion. **Conclusion**: Task assignment that promotes highly connected team communication networks may mitigate tacit knowledge loss in global software teams. Additionally, while members expressing openness to experience are likely to be particularly driven to perform, this is not entirely responsible for a global team's success.

## 1. INTRODUCTION

All but the smallest software development endeavors rely on the productive work of teams, whether co-located or (increasingly commonly) dispersed across locations. Prior research across numerous contexts has established that the intricacies of team dynamics may be revealed by studying members' communication[1] [1, 2], and that good communication is essential for building positive interpersonal relations in teams [3]. Among the specific assertions considered previously, research has revealed linkages between informal hierarchical communication structures and team performance for geographically distributed teams [4]. Team communication has also been linked to coordination efficiency [5] and to the quality of resultant software artifacts [6]. Thus, studying the details in and of team communication can provide valuable insights into the human processes involved during software development, including the importance of team members in communication structures, along with the reasons for, and consequences of, communication and coordination actions.

Similarly, aspects of team composition and team members' social and behavioral traits are also said to influence the outcomes of group-based activities. Such issues have been considered from multiple perspectives, including sociology and behavioral psychology relating to social identity [7], social capital [8] and personality psychology [9]. According to contemporary thinking, as well as observed practice in software development, individuals bring unique sets of knowledge[2] and skills to their collaboration during group work. These collective experiences (both prior and in-project), and in particular, those personal qualities that 'connect' during interactions, are influenced by participants' social and behavioral traits. Variations in these traits are said to determine if, and how, team members interact and the likelihood of teams being cohesive and productive [10].

Practitioners' communications and personalities and their effect on team members' behaviors and team output have therefore been receiving increasing attention in the software engineering research literature. For instance, Bird et al. [11] studied CVS records and mailing lists and concluded that the more software development an individual does the more coordination and controlling activities they must undertake. One of our own prior studies [12] found that just a few team members dominated project communication, and that these developers were crucial to their teams' organizational, intra-personal and inter-personal processes. Abreu and Premraj

---

[1] The terms "communication" and "interaction" are used interchangeably throughout this the paper to mean the exchange of information.

[2] Information or expertise acquired through previous experience or formal training.

[13] observed the Eclipse mailing list and found that increases in communication intensity coincided with higher numbers of bug-introducing changes, and that developers communicated most frequently at release and integration time. From a personality perspective, a study of 47 professional software engineers in ten Swedish software development companies found significant associations between personality factors and software engineers' behaviors [14]. Gorla and Lam's study [15] of the personalities of 92 high-performing IS professionals in Hong Kong uncovered that extroverted programmers outperformed those who were intuitive. Wang [16] also offered support for the linking of personality to team performance when reviewing 116 software project outputs.

While it has been observed that just a few members tend to dominate team communication during software development [17], little is known about these members' role in project knowledge diffusion. Knowledge diffusion is *the spread or transfer of knowledge from one part of a network (or individual) to another* [18]. Perhaps these highly active members communicate densely on specific tasks, and so they may be no more important in their team's overall knowledge diffusion process than those who communicate with fewer messages on comparatively higher numbers of software tasks. To this end, inferring practitioners' importance in knowledge diffusion based solely on the number of messages they contribute may be biased, resulting in unfulfilled expectations regarding practitioner performance and negative project consequences.

Similarly, there has been limited research focused on studying the potential influence of personality on members' involvement in knowledge diffusion during distributed and global software developments (GSDs). This is despite the belief that such work should help us to understand the potentially complex team dynamics in these environments [19]. Such explorations would seem to be particularly necessary given that these very teams are often challenged by reduced levels of awareness, group identification and shared understandings, due to team members' separation [20]. Thus, studying personality and behavioral issues in these settings, and understanding the potential impact of these variables on the performance of practitioners, should lead to recommendations that are likely to influence positive project outcomes [14].

In addressing these gaps we have led multiple explorations in an effort to contribute understandings around global team dynamics. For instance, we previously examined the role of core developers in global software teams to provide initial insights into the reasons for their exaggerated presence [12]. Additionally, we examined the potential influence of personality in global teams, providing insights into the profiles of Top Members [21]. In the current study we provide an extension to these works [12, 21], and bring together two distinct but related threads introduced above, communication and personality. We investigate the importance of active communicators in knowledge diffusion and the distribution of these practitioners' personalities as evident in language use, in order to provide insights into communication and personality variations among members in such a global setting. We mined the IBM Rational Jazz repository and used social network analysis (SNA) to cluster practitioners working across a set of teams into two groups (Top Members and Others, defined below). A number of SNA metrics (density, in-degree and closeness – refer to Section 3.2 for details) were then used to study practitioners' importance in knowledge diffusion, and these were triangulated with the exploration of particular linguistic usage. Finally, we performed further linguistic analysis to explore personality reflected in developers' messages, and related this evidence to records of activity in project history logs. We then relate the personality profiles of practitioners' to their involvement in knowledge diffusion. The findings from these activities are reported here.

This work makes multiple contributions. Firstly, we demonstrate that various analysis techniques may be systematically employed to deliver reliable and internally consistent results when studying human-related issues in empirical software engineering. Secondly, we extend previous work studying the communication and personality profiles of global developers. Finally, we provide recommendations for those tasked with leading globally distributed software development projects.

In the next section (Section 2) we present related work, and outline our specific research direction. We then describe our research setting in Section 3, introducing our procedures for data collection and measurement. In Section 4 we present our results, and discuss our findings. Section 5 then outlines implications of our results, and we identify potential threats to validity in Section 6. Finally, in Section 7 we draw conclusions.

## 2. RELATED WORK

We review related work in this section. In Section 2.1 we introduce research that has utilized repository data, and textual communications in particular, to explore human-centric aspects of software development processes. We then examine personality theories and models and how these have been applied to the study of software development practitioners' behaviors, in Section 2.2. Section 2.3 then addresses how personality may be studied from textual communications, with particular emphasis on the benefits of such analyses for global software development. Finally, we present our research questions in Section 2.4.

### 2.1 Analyses of Textual Communication

Software repositories and software history data have emerged as valuable sources of evidence of practitioners' interactions and communications [17, 22] (questions over data quality notwithstanding [12]). Accordingly, researchers have exploited process artefacts such as electronic messages, change request histories, bug logs and blogs to provide unique perspectives on the activities occurring during the software development process [1, 2]. In particular, previous work has focused heavily on studying communication patterns of software teams to explore and explain the knowledge diffusion process.

For instance, the Debian mailing list was used by Sowe et al. [23] to observe knowledge sharing among developers, with the authors finding that no specific individual dominated knowledge sharing activities in the Debian project. Crowston et al. [24] examined the work of the developers of five small open source software (OSS) projects using multiple explanatory approaches, including the principle of Bradford's law, and found that core groups of developers comprised only a small number of those contributing to the projects. Crowston and Howison's related study [25] found some OSS projects to be highly centralized (with just a few members communicating), and this pattern was especially pronounced for smaller projects. Additionally, it was revealed that most OSS projects had a hierarchical social structure, although there was greater communication modularity in larger projects [25]. Using the GTK+ and Evolution OSS projects, Shihab et al. [26] also established that only a small number of the developers participated in internet relay chat (IRC) meetings. Similarly, Shihab et al. [27] found communication activity to be correlated with software development activity when studying the GNOME project, where what was communicated was reflected in source code changes. Shihab et al. [27] observed that the most productive developers contributed 60% of the project's communication, and their interaction levels remained stable over project duration when compared to those of lesser contributing participants. Yu et al. [28] also found similar patterns of communication when studying artefacts' from the GNOME GTK+ project. Thus the weight of evidence suggests that OSS project teams tend to have relatively small groups of core developers who dominate the communication networks.

In more recent research investigating communication among multiple IBM Rational Jazz teams, Ehrlich and Cataldo [22] discovered that developers performed better when they occupied central positions in their team's communication network. However, their performance degraded when their networks were extended to multiple teams and across the entire Jazz project. Additionally, Ehrlich and Cataldo [22] revealed that those who were parties to dense network segments resolved more defects. Cataldo and Ehrlich [29] also studied IBM Rational Jazz teams' communications and revealed that teams that operated in a hierarchical communication structure completed more tasks in their iterations than those that worked in a small-world network communication structure. However, those that demonstrated the small-world communication pattern delivered higher quality software features.

These studies highlight the value obtained when investigating software practitioners' communication processes, especially in terms of how developers' behaviors in communication networks may affect or reflect their performance and that of their peers. In further considering developers' behavior we now review the relevance of personality theories.

## 2.2 Personality Theories and Models and Software Development

Personality theories support the use of validated tests to capture individuals' personality profiles. Two of the most frequently used and cited personality models are the Big Five [9] (with a variant called the Five Factor Model [30]) and the Myers-Briggs Type Indicator (MBTI) [31]. The Big Five personality model was developed from the theoretical stance that personality is encoded in natural language, and differences in personality may become apparent through linguistic variations [32]. In contrast, the MBTI model was developed deriving from the early work of Jung [33] on psychological types [31], with the underlying position that individuals evolve psychologically through experiences and this evolution shapes individuals' behaviors along different psychological types. Other notable instruments used for measuring personality include the Keirsey Temperament Sorter (KTS) model [34] and Cattell's 16PF model [35].

Of these models, the Big Five has emerged as the most dominant in assessing personality [36]. Additionally, although the MBTI is also widely used (e.g., [37]), psychologists have posited that the MBTI is suitable for assessing individuals' *self-awareness*, as against its common use in explaining performance [38]. Further, the Big Five personality profiles have been correlated previously with individuals' language use [39] – the phenomenon under consideration in this work. Therefore, a framework that uses the Big Five model was selected for use in this study.

As implied by its name, the Big Five personality model considers five personality profiles, being extroversion, agreeableness, conscientiousness, emotional stability (neuroticism) and openness to experience [40]. These profiles are revealed via the completion of a self-assessment questionnaire designed using Likert scale measures, known as the Big Five Factor Marker (BFFM). Extroversion describes individuals' desire to seek company and their drive for stimulation from the external world. Agreeable individuals are said to be cooperative, compassionate, and sensitive to others. Conscientiousness denotes a preference for order and goal-directed work. Individuals who are emotionally unstable (neurotic) have a tendency to show excessive negative emotions and anger. Finally, the openness to experience profile is associated with being insightful and open to new ideas.

Previous studies have identified correlations between personality, individual output and software team performance [16]. This latter stance is not universally supported, however, with others [41] finding expertise and task complexity to have greater influence on performance. Although there is some divergence over the view that personality impacts *performance*, there is strong evidence that personality does impact individual *behaviors* [14, 16]. These findings are highly relevant and have major implications for software development due to the creative and collaborative nature of development activities, and especially for distributed and global software efforts, where negative behaviors and a practitioner's willingness (or lack thereof) to engage and be part of knowledge diffusion may threaten collaboration. Thus, examining personality and its link to team members' engagement should help us to understand and more effectively manage the software development process as carried out by teams. We contend that the outcomes of such research would help us to make practice recommendations regarding team composition. The current objective of uncovering personality profiles from language use is, however, a non-trivial exercise in itself. In the following section (Section 2.3), we review theories examining the way personality profiles emerge in lexical examinations and we survey software development literature that demonstrates this method in use.

## 2.3 Textual Communication, Personality and Global Software Development

Evidence above concludes that individuals' behaviors are linked to their personality profiles. These personality traits can be detected in individuals' interactions, even if these occur in textual settings (e.g., email, blogs and text chat) [42]. For instance, in an experiment reported by Gill et al. [43] human judges were able to accurately rate subjects' personalities following a short examination of their written text. In particular, the extroversion profile was clearly evident from these linguistic observations. Support for linking personality to textual communication and word use has also been provided by Pennebaker and King [44].

Global software teams frequently use text-based tools (e.g., mailing lists, instant messengers and wikis) for requirements clarification, bug reporting, issue resolution and so on [28]. As noted in Section 2.1, these communications are often recorded in repository stores, which therefore have the potential to provide support for those studying team behaviors and engagements. In fact, data repositories and archives recording software developers' textual communication activities have already provided researchers with opportunities to study practitioners' social behaviors [45, 46]. However, while text analysis tools have been used previously to understand and predict various aspects of software development, only a few studies in this domain have considered examining personality from developers' textual communication. At the time of this review (which covered searches in the ACM Digital Library, IEEE Xplore, EI Compendex, Inspec, ScienceDirect and Google Scholar) studies were uncovered examining language use in relation to group member dominance [47], automatic personality recognition from speech [48] and personality perception in human agents [49]. However, only two studies (discussed below) were found that focused on analyzing personality from text [50, 51].

Given the growing attention directed to global and distributed software development, with many major market players such as Microsoft, IBM and Oracle using this approach in the delivery of major software releases [28], it seems timely to examine global software teams to understand the behavioral configurations under which these teams perform best. This would appear to be especially necessary and potentially useful given that GSD teams are hindered by distance, a factor that has been shown to obstruct effective coordination and control [20]. Given this impediment, issues related to personality imbalance are likely to have a negative impact on global teams' performance, especially given the way personality is said to affect team behaviors – negative incidences of which may affect both trust and team spirit, or members' willingness to engage and be part of their team's knowledge diffusion process [14]. This motivates the need for research efforts aimed at understanding this issue in the context of global teams.

### 2.4 Research Questions

*Communication*: Previous works reviewed above have found that a few developers from each team occupy the center of their team's communication network during software development [24, 27, 28], and there has been an increasing number of studies aimed at understanding why this pattern exists [12, 52]. However, little effort has been dedicated to the scrutiny of any differences in communication patterns across global teams of software practitioners, and particularly in terms of their potential role in knowledge diffusion throughout their team's communication network. For instance, in our own study of the differences between core developers and those who were less active, we found that while core developers were highly task- and achievement-focused, they were also largely responsible for maintaining positive team atmosphere [12]. However, these members' role in terms of knowledge diffusion – the transfer of knowledge to others in their communication networks – remains unexplored. While some practitioners no doubt communicate more than others during development, this may be related to their knowledge of specific software tasks under their portfolio. Thus, these core members may be no more important to knowledge diffusion than their colleagues who engage less, but on a wider cohort of features. We thus look to take a granular view of software practitioners' importance in knowledge diffusion in their team's communication networks by answering RQ1:

RQ1. How important are Top Members to their team's knowledge diffusion?

*Personality*: Although teams' personality configurations and their impact have been widely studied for collocated software development teams [14, 16], as noted above, Rigby and Hassan's work [50] is one of few that have assessed the personalities of distributed developers using textual communication. Their research applied a text analysis tool to identify Apache OSS developers' personality profiles from their mailing list exchanges. Their findings revealed that the top two developers were less extroverted than the other project members, and they scored lower on the openness to experience personality trait than the general population of contributors to the project. Additionally, the authors found that the top two developers scored similarly in terms of the conscientiousness personality profile. In contrast, Bazelli et al.'s examination of the StackOverflow forum found frequent contributors to be most extroverted [51]. Such divergent findings are insightful but also point to the need for further confirmatory research. Our goal was to build on these authors' work and study the personality profiles of software practitioners operating in a commercial (rather than OSS) development setting [21]. We therefore set out to address our second research question (RQ2):

RQ2. What are the personality profiles of Top Members?

*Communication and Personality*: The willingness of an individual to engage depends on their personal characteristics [53]. Personality studies have also shown that the actual nature of such engagements is driven by the behavior profiles of individuals [36, 37]. Thus, some individuals by their natural predisposition may be social and positive, while others may be particularly driven and outcome oriented, but still accommodating of others' opinions. In the same vein others may be more self-focused, maintaining only personal interest [10]. These variations in behaviors may be evident among global software developers, and may have implications for the overall performance of such teams. For instance, practitioners who are naturally inclined to collaborate and are social may occupy useful broker roles, facilitating knowledge diffusion across sub-teams and informing less communicative members. In a GSD environment where there are limited opportunities for engaging informally, such individuals could occupy very critical roles in the knowledge transfer process. Could there be a link then between personality profiles and members' willingness to be involved in knowledge diffusion? Are members' involvements in knowledge diffusion driven by their personality profiles? Establishing this linkage would be extremely useful for informing GSD team composition. We thus outline our third research question (RQ3) to guide this exploration:

RQ3. Do the personality profiles of Top Members differ from those of practitioners who are less active?

Finally, we proposed RQ4 to validate our findings against those provided by others around the way personality profiles are distributed in successful software teams [15]. While studying collocated teams researchers have observed specific configurations of personality distribution among successful teams [14, 54]. However, such evidence has not been provided for GSD teams. We thus assess the personality profiles of the distributed teams under consideration against those reported by previous work by answering our final research question:

RQ4. How are personality profiles distributed in a successful global team?

## 3. RESEARCH SETTING

We examined development artifacts from a specific release (1.0.1) of Jazz (based on the IBM$^R$ Rational$^R$ Team Concert$^{TM}$ (RTC)[3]), a fully functional environment for developing software and for

[3] IBM, the IBM logo, ibm.com, and Rational are trademarks or registered trademarks of International Business Machines Corporation in the United States, other countries, or both

managing the entire software development process [55]. The software includes features for work planning and traceability, software builds, code analysis, bug tracking and version control in one system. Changes to source code in the Jazz environment are permitted only as a consequence of a work item (WI) being created beforehand, such as a defect, a task or an enhancement request. Defects are actions related to bug fixing, whereas design documents, other documentation or support for the RTC online community are labeled as tasks. Enhancements relate to the provision of new functionality or the extension of system features. A history log is maintained for each WI. Team member communication and interaction around WIs are captured by Jazz's comment or message functionality. During development at IBM, project communication, the content explored in this study, was facilitated through the use of Jazz itself.

The release of the Jazz environment to which we were given access comprised a large volume of process data collected from distributed software development and management activities conducted across the USA, Canada and Europe. In Jazz each team has multiple individual roles, with a project leader responsible for the management and coordination of the activities undertaken by the team (and team members may also work across project teams – see Figure 1 to illustrate, which depicts the way individuals and teams are arranged in Jazz, in particular, note the common member shared by the Web Portal and Visual Studio Client teams in the intersection). All Jazz teams use the Eclipse-way methodology for guiding the software development process. This methodology outlines iteration cycles that are six to eight weeks in duration, comprising planning, development and stabilizing phases, and conforming generally to agile principles. Builds are executed after project iterations. All information for the software process is stored in a server repository, which is accessible through a web-based or Eclipse-based (RTC) client interface. The consolidated data storage and enforced project control mean that the data in Jazz is much more complete and representative of the software process than that in many OSS repositories. We provide details of our data extraction process and metrics definitions in the following two subsections (subsection 3.1 and subsection 3.2).

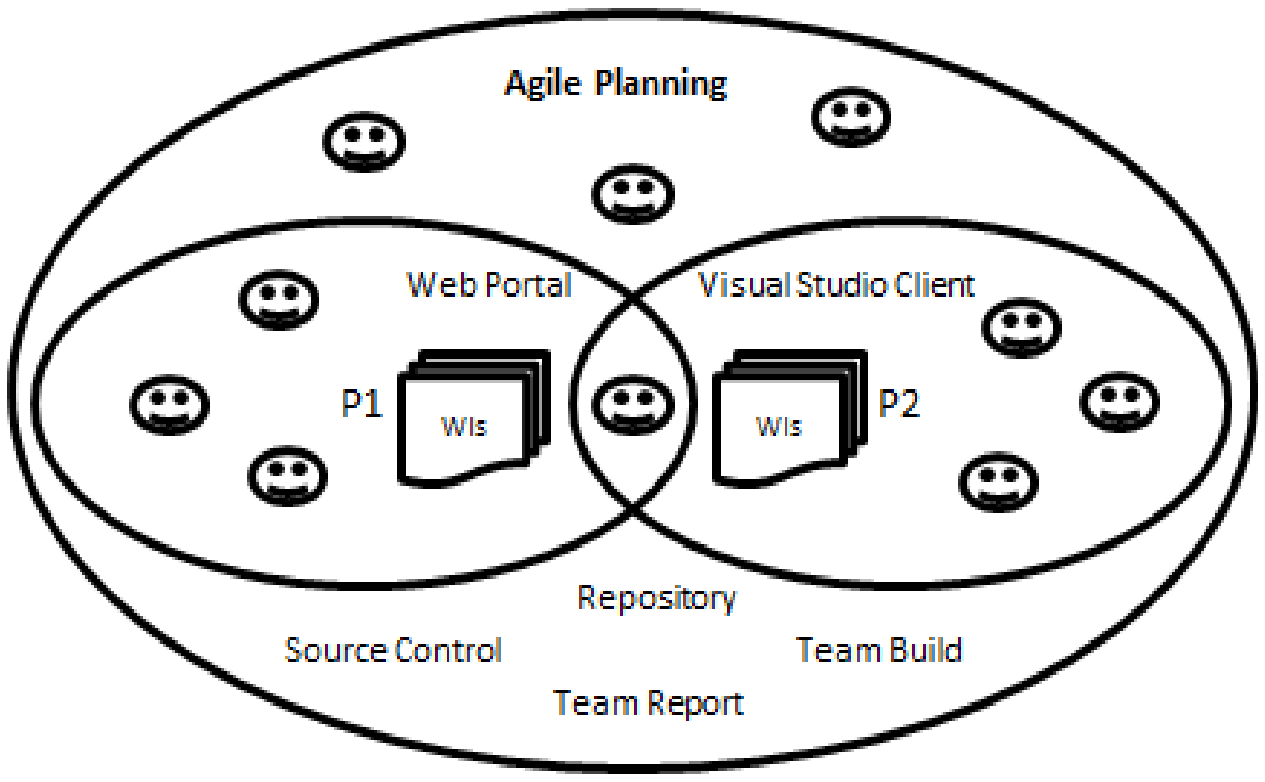

**Figure 1. Project teams' arrangement in Jazz**

## 3.1 Data Extraction and Preparation

Although a full explanation of data mining is beyond the scope of this paper we briefly report here the relevant steps performed in this project in terms of extracting, preparing and exploring the data under observation [56]. Data cleaning, integration and transformation techniques were utilized to maximize the representativeness of the data under consideration and to help with the assurance of data quality, while exploratory data analysis (EDA) techniques were employed to investigate data properties and to facilitate anomaly detection. Through these latter activities we were able to identify all records with inconsistent formats and data types, for example: an integer column with an empty cell. We wrote scripts to search for these inconsistent records and tagged those for deletion. This exercise allowed us to identify and delete 122 records (out of 36,672) that were of inconsistent format. We also wrote scripts that removed all HTML tags and foreign characters (as these would have confounded our analysis).

We leveraged the IBM Rational Jazz Client API to extract team information and development and communication artifacts from the Jazz repository. These included (in addition to the WIs discussed above):

- Project Workspace/Area – each Jazz team is assigned a workspace. The workspace (or project/team area P*n*) contains all the artifacts belonging to the specific team (see Figure 1 for a conceptual illustration).
- Contributors and Teams – a contributor is a practitioner contributing to one or more software features; multiple contributors in a Project Area form a team.
- Comments or Messages – communication around WIs is captured by Jazz's comment functionality. Messages ranged from as short as one word (e.g., "thanks"), up to 1055 words representing multiple pages of communication.

We extracted the relevant information from the repository and selected all the artifacts belonging to ten different project teams (out of 94) for analysis. The teams selected formed a purposive rather than random sample. Table 1 shows that the selected project areas represented both information-rich and information-rare cases in terms of WIs and messages. Project areas had tasks covering as few as two iterations to as many as 17 iterations (refer to Table 1), with varying levels of communication density. Density varies between 0 and 1 [57], where a task that attracted interaction from all the members in a team would have a density of 1, while those with no interaction would have a density of 0 (e.g., in terms of team members' density measures, a practitioner that communicated on 20 out of their team's 50 tasks would have a density of 20/50 = 0.4, denoted by D = c/n, where c is the number of tasks a member communicated on and n is the total number of tasks).

The selected project artifacts amounted to 1201 software development tasks comprising 692 defects, 295 support tasks and 214 enhancements, carried out by a total of 394 contributors (146 distinct members) working across the ten teams, with 5,563 messages exchanged around the 1,201 tasks. Although each of the ten teams had a slightly different balance in terms of role membership (refer to column four in Table 1), as the data were analyzed, we became increasingly confident that the cases selected were representative of those in the repository in terms of team members' engagement. We used SNA to initially explore the projects' communication from task-based social networks [58] resulting in a similar graph to that in Figure 2 for all of the ten project teams (note the dense communication segments for developers 12065 and 13664 respectively). Figure 2 represents a typical Jazz team's task-based directed social network, where network edges belonging to distinct contributors on individual software tasks are merged and color coded; edge color moved from red to brown (between one to five messages), brown to green (between six to ten messages) and then to a more pronounced green (ten or more messages). Network edges also increase in thickness corresponding to the number of messages that were communicated. The network vertices represented either a class image denoting a task or a contributor's unique identification number. In addition to the network visualizations (refer to Appendix A for all Jazz teams' (P1–P10) social networks), all ten project teams had similar profiles for network density (between 0.02 and 0.14) and closeness (between 0 and 0.06), as confirmed by formal statistical testing [59]. This consistency in SNA measures suggests that the teams selected were indeed relatively homogenous (refer to Licorish [59] for further details around our application of SNA to study IBM Rational Jazz software teams' collaboration in general).

**Table 1. Summary statistics for the selected Jazz project teams**

| Team ID | Task (Work Item) Count | Software Tasks (Project/ Team Area) | Total Contributors – Roles | Total Messages | Period (days) – Iterations |
|---|---|---|---|---|---|
| P1 | 54 | User Experience – tasks related to UI development | 33 – 18 programmers, 11 team leads, 2 project managers, 1 admin, 1 multiple roles | 460 | 304 - 04 |
| P2 | 112 | User Experience – tasks related to UI development | 47 – 24 programmers, 14 team leads, 2 project managers, 1 admin, 6 multiple roles | 975 | 630 - 11 |
| P3 | 30 | Documentation – tasks related to Web portal documentation | 29 – 12 programmers, 10 team leads, 4 project managers, 1 admin, 2 multiple roles | 158 | 59 - 02 |
| P4 | 214 | Code (Functionality) – tasks related to development of application middleware | 39 – 20 programmers, 11 team leads, 2 project managers, 2 admins, 4 multiple roles | 883 | 539 - 06 |
| P5 | 122 | Code (Functionality) – tasks related to development of application middleware | 48 – 23 programmers, 14 team leads, 4 project managers, 1 admin, 6 multiple roles | 539 | 1014 - 17 |
| P6 | 111 | Code (Functionality) – tasks related to development of application middleware | 25 – 11 programmers, 9 team leads, 2 project managers, 3 multiple roles | 553 | 224 - 13 |
| P7 | 91 | Code (Functionality) – tasks related to development of application middleware | 16 – 6 programmers, 7 team leads, 1 project manager, 1 admin, 1 multiple roles | 489 | 360 - 11 |
| P8 | 210 | Project Management – tasks under the project managers' control | 90 – 29 programmers, 24 team leads, 6 project managers, 2 admins, 29 multiple roles | 612 | 660 - 16 |
| P9 | 50 | Code (Functionality) – tasks related to development of application middleware | 19 – 10 programmers, 3 team leads, 4 project managers, 2 multiple roles | 254 | 390 - 10 |
| P10 | 207 | Code (Functionality) – tasks related to development of application middleware | 48 – 22 programmers, 12 team leads, 2 project managers, 1 admin, 11 multiple roles | 640 | 520 - 11 |
| ∑ | **1,201** | | **394 contributors,** comprising 175 programmers, 115 team leads, 29 project managers, 10 admins, 65 multiple roles | **5,563** | |

## 3.2 Description of Measures

We define our measures in this subsection. First, we outline how our dataset was partitioned to study the issues under consideration. We then define the measures that were used for studying knowledge diffusion. Finally, we explain how personality was operationalized.

*Identifying Top Members and Others (via their engagements)*: We first created a baseline level of developer contribution using an approach similar to that used by Crowston et al. [24], and selected all practitioners whose communication density measure was greater than or equal to 0.33 (i.e., they communicated on a third or more of their teams' project tasks – density is denoted more generally as density = c / (n(n-1)/2), where c is the number of messages communicated and n is the total number of members present in the communication network), and labeled them as *Top Members* (see our earlier work [12] for further details). Fourteen contributors across the ten project teams met this selection criterion – a smaller number than preferred in terms of supporting our intended analysis. We therefore relaxed our initial requirements and selected the top two communicators of each team, following the approach used by Rigby and Hassan [50], which increased the total number of top members by six (to mean 20 practitioners in total but comprising 15 distinct developers). The other members were then placed in the group labeled *Others*. Eight of the 15 distinct Top Members were programmers, five were team leaders and two were project managers [12]. As a second step to validating that these practitioners were indeed Top Members, analysis confirmed that these individuals had initiated more than 41% of all software tasks, made more than 69% of the changes to these tasks and resolved nearly 75% of all software tasks undertaken by their teams (refer to [12] for further details). On the basis that the selected practitioners had the highest levels of *engagement* in team *communication* and *task changes* we believe that these members can indeed be considered to be core developers – those most engaged and active in their teams' performance [12]. We used these groups (*Top Members* and *Others*) as our unit of analysis, but also considered our findings more generally at the individual, team and project levels.

*Measuring Knowledge Diffusion (via SNA metrics)*: Social Network Analysis (SNA) is used to quantify aspects of network structures in order to support pattern identification in communication networks [60]. This technique employs mathematical analysis and pictorial representations of the patterns of interaction and relationships among individuals – and potentially other components – during group processes [61]. Concepts such as cohesion, equivalence, power and brokerage are used to explain the characteristics of network actors [57]. Of these concepts, the most important mathematical measurement for SNA is cohesion, measured by density and centrality. Density provides an overall measurement of the connectedness of the network [57], whereas centrality (also called degree or degree centrality) denotes the level of individual interaction [62]. Visualization of interaction networks, also called sociograms, is often used for uncovering collaboration patterns and the flow of information that may not be so evident from numerical values [62]. In these visualizations, individuals are represented by nodes, and their associations are illustrated through lines that connect these nodes. An examination of a sociogram will quickly unveil who is communicating (or not), who is most central to the team, which members are acting as hubs or brokers, and so on – refer to Figure 2 for illustration, which demonstrates a typical Jazz team's task-based directed social network (described above).

SNA has been shown to have value in many domains, including security [63], political science [64], education and communication [65], as well as in software engineering [1, 58]. While it is generally recommended that caution be shown when explaining the consequences of social network patterns due to the many reasons people may interact [66] – whether the reason be due to peer authority, social pollution or otherwise – collaboration in professional software engineering settings is generally linked to work task execution, and active collaboration has been shown to have a positive impact on team productivity [67, 68]. Thus, SNA techniques provided utility for this work, and were used to study knowledge diffusion (in addition to separating practitioners, as presented above). We use multiple SNA measures to provide a measurement of knowledge diffusion, including density, in-degree and closeness, as considered below.

Individuals involved in highly *dense* communication network segments have been shown to dominate coordination and collective action [69], and are seen as significant to their teams' knowledge diffusion process [70]. In defining our density measure above, we noted that if a practitioner communicated on 20 out of their team's 50 tasks, that member would have a density of 20/50 = 0.4, similarly a member that communicated on 40 of the 50 tasks would have a density of 40/50 = 0.8. The more a member communicates the more they are considered to be involved in knowledge exchange.

Similarly, *centrality* (introduced above) provides an assessment or rank of network ties and social prestige. Local centrality reflects immediate connections to those around (in-degree and out-degree[4], defined formally by degree (node_i) $=\sum_j X_{ij}$, $= \sum_j X_{ji}$), while global centrality refers to individuals' closeness[5] (interconnectivity or cohesiveness) to many others in the wider network [71]. Closeness is measured formally as: Closeness (node_x) = (n-1) * $C_C(x)$, where $C_C(x) = 1 / \sum y{\in}U^{d(x,y)}$, where d(x,y) is the length of the shortest path between vertices x and y (the theoretic distance), and *U* is the set of all vertices [17]. Very *central* and *close* network nodes are considered to play an important role in maintaining both local and global information transfer and network connectivity, and by extension, knowledge diffusion. Such nodes are also generally regarded as relatively powerful [72].

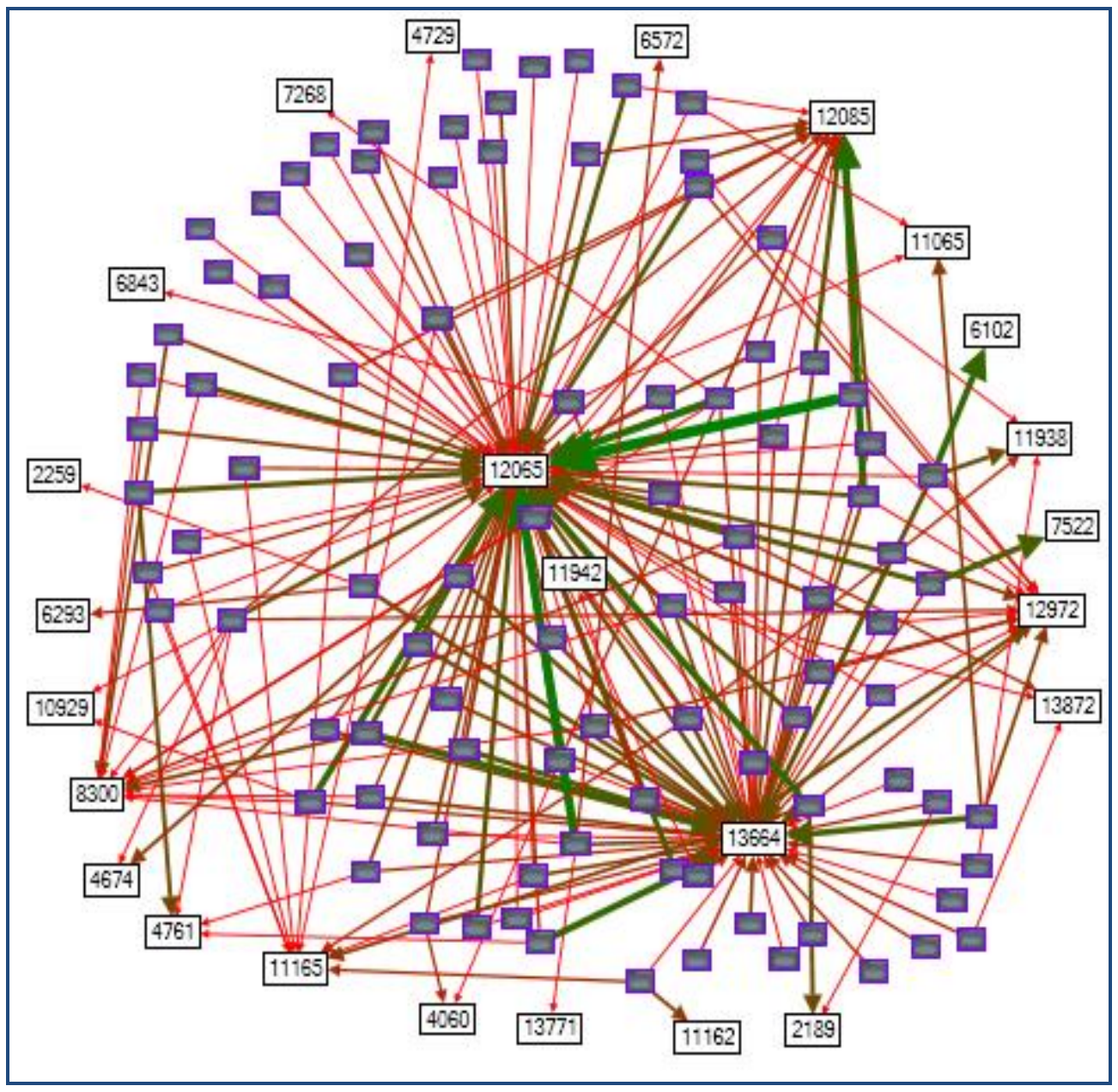

**Figure 2. A Jazz team's network graph showing highly dense network segments for contributors "12065" and "13664"**

To illustrate the above concepts, consider Figure 3 which shows a simple knowledge network with four different network segments (a, b, c, and d). In Figure 3(a) the member signified by a solid blue node has a degree centrality of 6, i.e., (s)he communicated with 6 other individuals. When compared to the other individuals in this network segment (whose degree centrality ranges between 2 and 3), this is the most central member in this group of communicators. In Figure 3(b) the solid green node (member) forms the knowledge link (hub or bridge) between two network segments (Figure 3(a) and Figure 3(c)), while in Figure 3(c) the solid red node, although playing the role of bridge for segments Figure 3(c) and Figure 3(d), has a degree centrality of 1 in Figure 3(c), and is the weakest communicator in both Figure 3(c) and Figure 3(d). The most dense network segment is seen in Figure 3(d) where all members have a degree centrality of between 5 and 6 (with 6 being the highest possible degree centrality). In this network segment (Figure 3(d)), the solid black member has a density of 1 (the maximum value for the density measurement).

---

[4] In-degree denotes the number of connections (edges) that point towards a node, while out-degree is the number of connections originating (pointing outwards) from that node.

[5] Closeness measures the shortest distance between nodes, in terms of both direct and indirect connections, so, the lower the closeness measure for a given node the more reachable that node is to the other members in the network.

Similarly, overall, Figure 3's closeness measure would be 0, meaning all network nodes are interconnected (reachable).

Each of the nodes in Figure 3 would occupy a unique role in local and global knowledge diffusion around the different segments of the network. For instance, while the solid blue node in Figure 3(a) is likely to play a significant role in spreading (duplicating) knowledge across this local network segment, this knowledge would only remain local in the absence of the other members' connections to Figure 3(b) – the green node. The green node in Figure 3(b) is playing a very important (close) role in filling a structural hole[6], and providing the additive (global) knowledge link to Figure 3(c), and by extension, Figure 3(d).

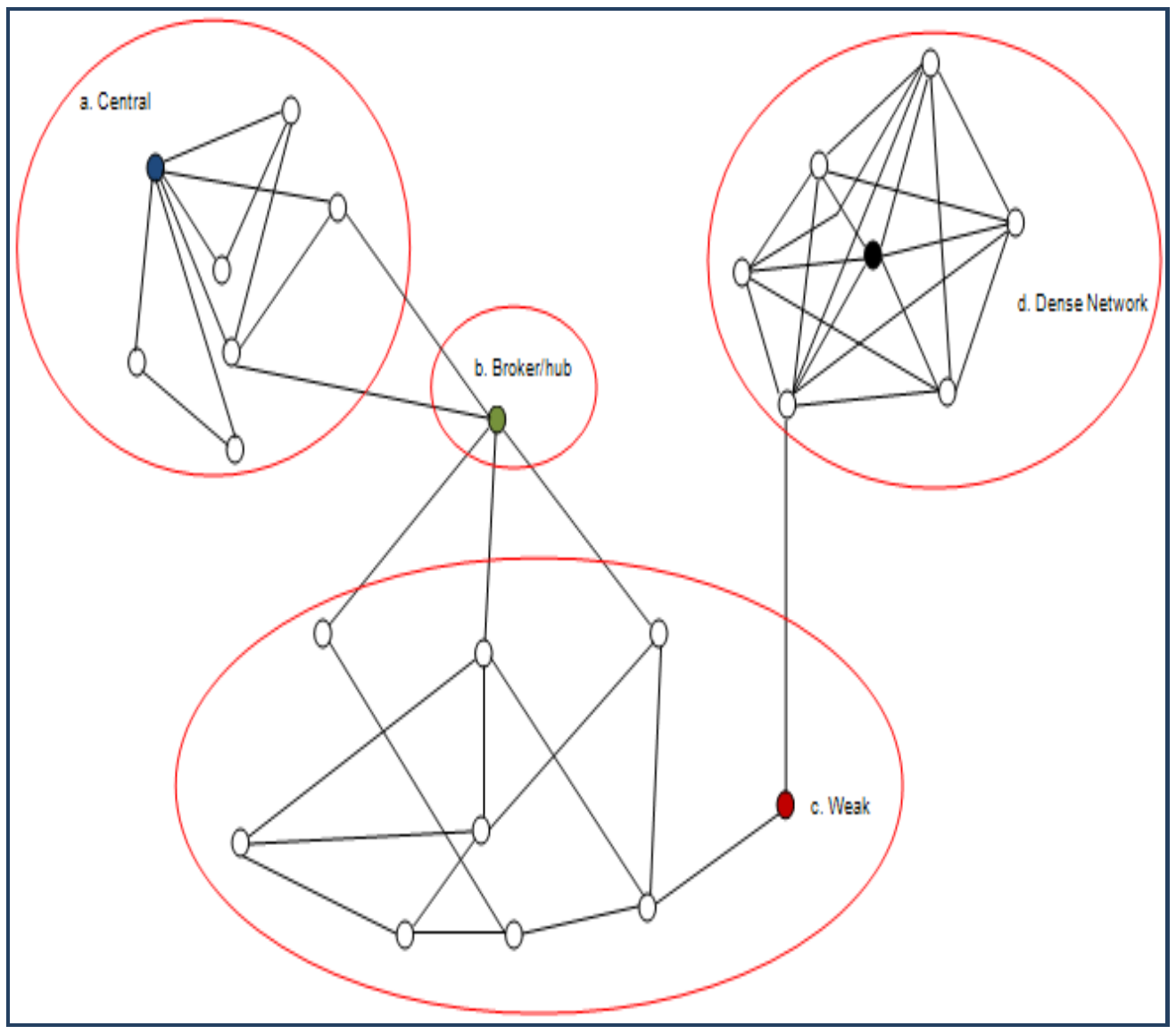

**Figure 3. Communication network highlighting interaction patterns of team members**

As noted above, we used *density* and *centrality* measures for *in-degree* and *closeness* to study the way practitioners were *involved with knowledge diffusion*. These measures have also been used previously in other contexts to study this issue (see [73], for example). In adopting these measures we are able to assess both the frequency with which practitioners shared knowledge in local segments of the network and their reach across the wider knowledge networks. We assess those communicating on the same tasks as being equally aware of the specific task's knowledge, which may be conveyed to others through further tasks' connections (e.g., in the actual Jazz communication network in Figure 2 contributors 11165 and 4060 are expected to have common knowledge of the tasks they share, which may be conveyed to contributor 6293 via contributors' 11165 and 6293 shared task). Finally, we perused teams' sociograms and also considered how practitioners used work and achievement terms during their exchanges to triangulate our SNA metrics (refer to [45, 46, 74, 75] for examples of our application of psycholinguistics to the study of software practitioners' discourses).

*Measuring personality (via linguistic patterns)*: Previous studies examining personality profiles from textual communication have successfully employed dictionary-based approaches (for example: [50, 76]). Of these approaches, research has found the linguistic inquiry and word count (LIWC) scales to accurately support personality assessment from communication [77]. This approach has also found wider support in terms of being linked to personality trait assessment in the psycholinguistic literature when compared to others [39]. Consequently, this study employed the LIWC tool in order to ascertain software practitioners' personality profiles from their written communication.

The LIWC software tool was created after four decades of research using data collected across the USA, Canada and New Zealand [44]. Data collected in creating the LIWC tool included all forms of writing and normal conversations. This tool captures over 86% of the words used during conversations (around 4,500 words). Words are counted and grouped against specific types, such as negative emotion, social words, and so on. Written text is submitted as input to the tool in a file that is then processed and summarized based on the LIWC tool's dictionary. Each word in the file is searched for in the dictionary, and specific type counts are incremented based on the associated word category. The tool's output data include the percentage of words captured by the dictionary, standard linguistic dimensions (which include pronouns and auxiliary verbs), psychological categories and function words (e.g., negative, social) and personal dimensions (e.g., work and leisure). These different dimensions are said to capture the psychological profiles of individuals by assessing the words they use [44].

In terms of our *measurement of personality*, neuroticism has been found to be associated with frequent use of first-person pronoun and negative emotion words, while being negatively correlated with the use of positive emotion words and articles. Extroversion was found to be correlated with positive emotion words and social process words, but had lesser association with negations, tentativeness and the expression of negative emotion. People possessing the conscientiousness profile were found to use more positive emotion words and few negation, discrepancy and negative emotion words. People characterized by the openness to experience profile communicated with more articles, longer words and tentativeness, but these individuals rarely used self-references and past tense words. Finally, those who were more agreeable used more positive emotion and self-references, but fewer articles and less negative emotion. Similar to Rigby and Hassan [50], we used the LIWC tool to analyze practitioners' language to provide an indication of their personality profiles. In line with these authors, we utilized composite measures formulated by aggregating the LIWC factors that correlated with the *Five-Factor* scores of Pennebaker and King [44]. For example, openness to experience was formed using the composition (articles (e.g., a, an, the) + long words (words > 6 letters) – first person pronouns (e.g., I, me, mine) – present tense words (e.g., is does, here) + exclusive words (e.g., but, without, exclude) + tentative words (e.g., maybe, perhaps, guess) + insightful words (e.g., think, know, consider) + causation words (e.g., because, effect, hence)).

Whereas Rigby and Hassan [50] looked at the behaviors evident in the language of the top committers compared to other contributors in the Apache OSS project, we considered language use and behaviors of the Top Members and Others for ten Jazz project teams, and also examine personality at the team level more generally. We also evaluate our personality findings in our exploration of practitioners' communication during teamwork. Furthermore, given that we were mining data from a commercial repository where challenges related to contributors' multiple

[6] The existence of non-redundant information in two separate network clusters.

email addresses and aliases and identifying real project contributors [78] are less likely to occur, we expected to provide significant and robust contributions regarding the communication and personality profiles of successful global teams.

Finally, relevant *software project success* indicators include measures related to software impact on the development organization, the reviews of post-release customers, and actual software usage [112]. Thus, given the impact IBM Rational products (included in the Jazz repository) have had on IBM and many other organizations, with over 30,000 companies using these tools, and that these products have been positively reviewed and tested by those companies, it is contended here that Jazz teams are successful (see http://www.jazz.net for details).

# 4. RESULTS AND ANALYSIS

We applied multiple analysis techniques (noted in Section 3.2) and formal statistical testing to the pre-processed data (refer to Section 3.1), and report our results and analysis in this section. We first examine the results for Top Members' knowledge diffusion in Section 4.1, comparing measures for Top Members against those of their colleagues to answer RQ1. We next report the results and provide discussions for each of the 15 Top Members' personality profiles (refer to Section 4.2) in order to answer RQ2. In answering RQ3 we inspect the personality profiles of the less active members, and compare their results to those of the Top Members in Section 4.3. Finally, in Section 4.4 we provide the confirmatory results and discussions for our examination of personality profiles at the team level in order to answer RQ4.

## 4.1 Top Members' Knowledge Diffusion

We provide the results of our SNA in Table 2, and also make comparisons between Top Members and their colleagues. Table 2 shows that Top Members were actively involved in their teams' communication, and played a key role in their teams' knowledge diffusion. In terms of in-degree (refer to the third column of Table 2), Top Members contributed over 62% of the measures for teams P6 and P7, and Top Members on P1, P2, P9 and P10 contributed a combined 51.1%, 46.6%, 44.7% and 44.7% of their teams' measures, respectively. Overall, the Top Members had a mean in-degree score of 51.9%. This number represents 21.8% of the overall measure for all project areas. The density figures in Table 2 show a similar pattern. Here it is revealed that contributors 4661 (of P1), 12972 (of P7) and 12065 (of P6) had density measures of 0.85, 0.80 and 0.74 respectively, and the overall mean density measure for Top Members was 0.48, compared to the mean project teams' (P1 – P10) density measure of 0.07 (i.e., Top Members communicated on 48% of the tasks compared to their overall teams' score of 7%). These measures were compared for statistically significant differences. When the density scores were checked for normality it was noted that there was no violation of the normality assumption [79]. A Levene's test for equality of variance revealed unequal variances for the two groups (Top Members and their teams) ($p < 0.001$). Thus, the parametric independent sample *t*-test was conducted to test the mean density measures for significant differences. This revealed a statistically significant difference between the density of Top Members and those of their teams' ($p < 0.001$). This result show Top Members communicated on nearly seven times as many software tasks as their teammates, and were significantly more likely to disseminate more knowledge across their teams' communication network.

In considering the closeness results in Table 2 (shown in the last column), a slightly different pattern is observed. Noticeably, the network measures for closeness for Top Members are not much lower than those for the teams' networks (with an overall mean score of 0.00 for Top Members and 0.01 for the teams). In fact, Table 2 shows that for project area P7 the closeness measure for the team's network is lower than those of the Top Members (0.00 versus 0.01 for each Top Member). The closeness scores were checked for statistically significant differences. Firstly, an examination of the standardized skewness and kurtotis coefficients for the closeness measures revealed serious departures from normality [79]. Thus, a non-parametric Mann-Whitney U test was used to compare the scores of the Top Members and those of their teams' (refer to Table 2). This test did not reveal any statistically significant difference in the closeness scores for Top Members and their teams ($p = 0.404$).

Figure 4 is used to triangulate this finding through visual assessments. We removed all the connections (messages) of the two Top Members (developers 12065 and 13664 respectively) in Figure 2 thereby deriving Figure 4. This latter figure (Figure 4) shows that although Top Members communicated exclusively on nearly a third of the software tasks (refer to class images without connections), when these members were removed from their team's networks, the Other contributors still remained close (i.e., reachable whether directly or via their sharing in other tasks' knowledge networks). In addition, Figure 4 reveals that most of the Other members did not communicate densely on many software tasks. These findings denote that, overall, IBM Rational Jazz members were all very accessible irrespective of their levels of active contribution.

We further triangulate our results above by examining linguistic variations for work and achievement forms of language use, as an indicator of the drive to share expertise. We examined word usage that has been shown to indicate interest in work and achievement [44, 80]. For example, those words clustered under the work dimension include "feedback", "goal", "boss", "overtime", "program", "delegate", "duty" and "meeting", while words including "accomplish", "attain", "closure", "resolve", "obtain", "finalize", "fulfill", "overcome" and "solve" have been shown to signal interest in task achievement (refer to the LIWC dictionary source for the full corpus of words considered under these categories here: http://www.liwc.net/).

All communication from the two groups were aggregated, and then tokenized. Overall, those practitioners classified as Top Members communicated 2567 messages while the Others communicated 2996 messages. Given the skweness of contributions we normalized our measures by examining percentages of work and achievement linguistic usage. We first examined our data distributions for normality using the Kolmogorov-Smirnov test which revealed violations of the normality assumption. We thus examined work and achievement linguistic usage between Top Members and Others for significant differences using the non-parametric Mann-Whitney U test. These results show that Top Members used more work (Mean Rank, Top Members = 2841.84, Others = 2742.06, $p = 0.013$) and achievement (Mean Rank, Top Members = 2828.21, Others = 2753.68, $p = 0.063$) utterances than the Other practitioners. We consider these results in relation to previous theories next.

**Table 2. Social network measures for top members and their teams (P1 – P10)**

| Team ID | Contributor | Top Members' In-degree | | Density | | Closeness | |
|---|---|---|---|---|---|---|---|
| | | In-degree | (% of team's measure) | top | team | top | team |
| P1 | 4661 | 46 | 32.6 | 0.85 | 0.08 | 0.01 | 0.01 |
| | 2419 | 26 | 18.4 | 0.48 | | 0.01 | |
| P2 | 4661 | 83 | 33.3 | 0.74 | 0.05 | 0.00 | 0.00 |
| | 2419 | 33 | 13.3 | 0.29 | | 0.00 | |
| P3 | 13722 | 15 | 22.1 | 0.50 | 0.08 | 0.01 | 0.06 |
| | 4674 | 7 | 10.3 | 0.23 | | 0.01 | |
| P4 | 13740 | 85 | 19.4 | 0.40 | 0.05 | 0.00 | 0.01 |
| | 11643 | 70 | 16.0 | 0.33 | | 0.00 | |
| .P5 | 4749 | 55 | 18.6 | 0.45 | 0.05 | 0.00 | 0.01 |
| | 4674 | 39 | 13.2 | 0.32 | | 0.00 | |
| P6 | 12065 | 82 | 35.7 | 0.74 | 0.08 | 0.01 | 0.00 |
| | 13664 | 61 | 26.5 | 0.55 | | 0.00 | |
| P7 | 12972 | 73 | 35.1 | 0.80 | 0.14 | 0.01 | 0.00 |
| | 13664 | 57 | 27.4 | 0.63 | | 0.01 | |
| P8 | 12702 | 59 | 15.8 | 0.28 | 0.02 | 0.00 | 0.02 |
| | 2102 | 33 | 8.8 | 0.16 | | 0.00 | |
| P9 | 6572 | 29 | 25.4 | 0.58 | 0.12 | 0.01 | 0.01 |
| | 12889 | 22 | 19.3 | 0.44 | | 0.01 | |
| P10 | 6262 | 127 | 34.8 | 0.61 | 0.04 | 0.00 | 0.00 |
| | 13722 | 36 | 9.9 | 0.17 | | 0.00 | |
| **Mean** | **-** | **51.9** | **21.8** | **0.48** | **0.07** | **0.00** | **0.01** |

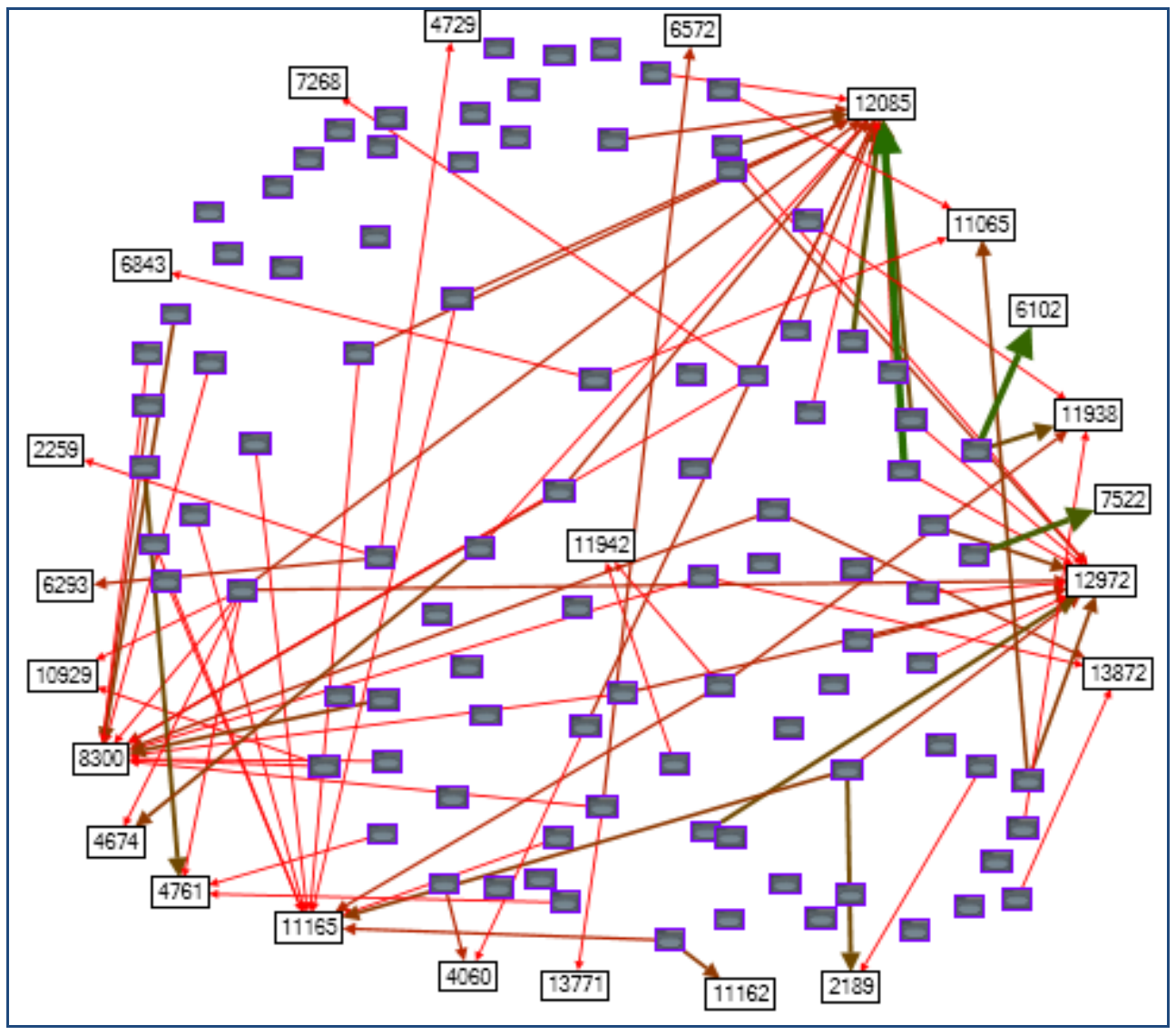

**Figure 4. A Jazz team's network graph showing Top Members ("12065" and "13664") and their connections removed**

*Discussion of RQ1. How important are Top Members to their team's knowledge diffusion?* Our findings in this study suggest that Top Members occupied an integral role in their team's knowledge diffusion process. We found that, apart from the higher frequency with which Top Members communicated [12], there was also substantial disparity between the numbers of tasks Top Members communicated on compared to the rest of their team members (refer to Section 4.1, and Appendix A – note here that all ten project teams had extremely dense communication segments around Top Members). The results show that Top Members communicated in relation to seven times as many tasks as their teammates, on average. These findings show that, beyond the incidence (or volume) of messages, Top Members also maintained interest on many tasks, suggesting that they may have been integral in disseminating knowledge to both local and global segments of the network. The level of interest shown by Top Members is indeed interesting when considering that distance mediated these members, and thus, likely delayed the development of team trust and the ability to rapidly form relationships [81, 82].

Of course it is only through the actual communication of messages that social networks are formed, and team norms evolve, making teams functional [83, 84], after divergent opinions become homogenous. Previous work suggests that organizations tend to benefit from the social and personal relationships of individual employees, which has been shown to encourage knowledge diffusion [85]. The knowledge that is transferred (whether structured or unstructured) also mediates the effect of knowledge diffusion. The study of network structures through various measures, although not necessarily revealing the *nature* of the knowledge being transferred, provides a starting point to understanding the way individuals and teams communicate and transfer knowledge. Previous works in the organization science space have indeed shown the relevance of these techniques for dissecting actors' involvement in knowledge diffusion [86].

Central actors have particularly been held to communicate because of how much they know. Research has shown that these members are critical to their teams' shared understanding [24, 87], and early works investigating the significance of centralized group members during teamwork have stressed the importance of these actors to their team's performance [88, 89], and particularly for their teams' knowledge diffusion. In fact, while the SNA metrics for density, in-degree and closeness used to investigate knowledge diffusion in this study may be assessed as reflecting Top Members' occupancy of *only* structurally important roles (as depicted in the sociograms in Appendix A), rather than intellectual or social status, our deeper analysis of the *actual content* of Top Members' messages confirmed that their messages comprised largely of task-based information, ideas and comments, scaffolding and instructions. Furthermore, the incidence of these forms of discourse was much higher for Top Members when compared to those less active [52]. Here in this work our analysis of Top Members' linguistic usage also confirms their significance to knowledge diffusion, beyond their structural roles.

In addition, an individual's linguistic usage has been shown to reflect the realities of their situation, their mood, and behavioral traits [43, 90, 91]. Thus, our observation for the higher level of work (e.g., feedback, goal, delegate) and achievement (e.g., accomplish, attain, resolve) linguistic usage by Top Members suggests that they were central to such processes in their teams. Beyond knowledge diffusion, such an orientation would have an impact on new members joining these teams, where establishing social ties with work- and achievement-orientated members (Top Members) could no doubt influence such new members' desire to be similarly oriented. This proposition is particularly plausible given Top Members' structural dominance in their teams' communication networks. In fact, our previous contextual examination of six Jazz members' communication confirmed that they also occupied interpersonal roles [12], which may also be contagious for other new members.

Extending our previous evidence [12], our in-depth exploration of one of the Jazz teams' networks here revealed that Top Members communicated exclusively on nearly a third of the software tasks that were performed by their team (refer to Figure 4 for network nodes with no edges). This finding coincides with our evidence for these members' very high level of involvement in actual software development tasks, for all ten teams [12]. This could be detrimental for retaining teams' tacit knowledge, should Top Members leave their teams [92], which would likely create structural holes. Those occupying structural holes in their social networks, as noted for Top Members, have been shown to hold positions of power, as they often possess additive as against redundant knowledge [93]. These members also tend to access information from many diverse network clusters, suggesting that Top Members in our study benefited from their access to this diversity of views on the tasks in which they participated.

Of note, however, is our SNA findings for Jazz teams' highly connected networks. We observed that Top Members' *closeness* measures, although lower than those of their *Other* teammates (and signaling that they were more connected), were not significantly different. In fact, in Figure 4 we observed that even when the communications for Top Members were removed from their team's social network, the network still remained highly connected (notwithstanding the tasks that Top Members communicated on exclusively). Less active members were connected through their involvement on software tasks, whether or not Top Members were present. Although it is unclear whether this is a deliberate strategy employed by Jazz teams to promote knowledge redundancy, or whether this evidence reflects the process of self-organization among high-performing globally distributed agile software practitioners, this is a positive result with regard to knowledge diffusion and retention; and more generally, for cross training [94] in globally distributed software development contexts. This pattern would have a positive effect on these teams retaining tacit knowledge should Top Members leave. Perhaps it would also benefit their team if similarly oriented individuals were hired after their departure. We examine this issue further in the next section, specifically assessing the personality profiles of Top Members.

## 4.2 Top Members' Personality Profiles

From the results shown in Table 3 it is evident that openness to experience (see the higher values in column four), extroversion (column three) and agreeableness (column five) were the most pronounced personality profiles exhibited by the Top Members. We checked to see if there were significant differences in the predominance of these personality profiles, first using Shapiro-Wilk tests to evaluate normality. These tests revealed that the measures for the five personality profiles were all normally distributed. We therefore conducted a two-way ANOVA test which confirmed that there were statistically significant differences ($p < 0.01$) in the predominance of the five personality profiles for Top Members. Given this finding a series of *t*-tests was conducted at the Bonferroni adjusted level of 0.005 (i.e., 0.05 divided by 10 analyses). These results confirmed that Top Members' openness to experience profiles were significantly higher ($p < 0.005$) than all other profiles apart from extroversion. We also noted that profiles for extroversion and agreeableness were much higher than those for neuroticism and conscientiousness, respectively ($p < 0.005$). Table 4 also shows that the top practitioners in our sample used similar linguistic processes across project teams (see Table 4 for *t*-test results comparing differences in three randomly selected linguistic categories for Top Members involved in multiple teams). Overall, 11 of the 15 Top Members across the ten teams exhibited pronounced amounts of the openness to experience personality profile (see Open measures for contributors 12889, 6262 and 12702 in Table 3, for example), while this profile was lower for other Top Members (see the Open measure for contributor 12972 in Table 3, for example). Evidence of extroversion was also very apparent among these practitioners, and particularly for practitioners 4661, 2419, 13722 and 4674; refer to the Extro measures for these members in Table 3. There was also evidence of the agreeableness profile (see Agree measures for 12889, 11643 and 6262, for example). Profiles of neuroticism and conscientiousness were also evident but less pronounced in the discourses of other Top Members (see Neuro and Consc measures for 11643, 12889 and 2102 in Table 3). Overall, the conscientiousness profile was the least evident among the Top Members considered in our sample (refer to measures for the Consc column in Table 3).

We anticipated that message volume may have affected the pattern of results obtained; however, this was not the case. Additionally, we did not find any linkage between the role(s) occupied by each individual and communication and task engagements. In fact, given the high number of programmers in the Top Members group we suspect that the Jazz collaboration environment required that programmers communicate more than might normally be expected of someone in that role. In Jazz, a person occupying the formal "Programmer" (contributor) role is defined as a contributor to the architecture and code of a component, the "Team Leader" (component lead) is responsible for planning and for the architectural integrity of the component, and the "Project Manager" (PMC) is a member of the project management committee overseeing the Jazz project.

**Table 3. Top Members' personality scores (values are relative not absolute)**

| Contributor | Neuro | Extro | Open | Agree | Consc |
|---|---|---|---|---|---|
| 4661 | -7.24 | 61.89 | 42.73 | 28.86 | 1.84 |
| 2419 | -46.88 | 73.23 | 38.21 | 15.30 | 3.63 |
| 13722 | 12.71 | 49.68 | 37.39 | 27.05 | 5.19 |
| 4674 | 15.77 | 51.96 | 43.88 | 34.22 | 2.79 |
| 13740 | 10.79 | 40.66 | 46.56 | 35.73 | 8.10 |
| 11643 | 37.94 | 27.91 | 48.72 | 45.66 | 6.63 |
| 4749 | 36.95 | 27.00 | 44.40 | 41.59 | 7.80 |
| 12065 | 18.72 | 33.26 | 39.29 | 36.09 | 6.18 |
| 12972 | 33.03 | 27.05 | 31.67 | 39.70 | 5.15 |
| 13664 | 23.64 | 36.86 | 53.58 | 33.20 | 7.42 |
| 12702 | 12.27 | 44.52 | 55.01 | 39.85 | 2.56 |
| 2102 | 17.17 | 44.60 | 53.79 | 34.93 | 9.93 |
| 6572 | 34.30 | 28.00 | 46.72 | 40.99 | 8.25 |
| 12889 | 39.84 | 36.12 | 57.17 | 49.99 | 1.60 |
| 6262 | 31.50 | 32.19 | 56.12 | 42.87 | 8.35 |

We provide a form of triangulation for our findings above by examining Top Members' actual usage pattern of positive and social process words (e.g., "give", "buddy", "love", "explain", "friend", "beautiful", "relax", "perfect", "glad" and "proud") and complex words (e.g., words with more than six characters in length or those with multiple syllables), linguistic dimensions that contribute to the measurement of extroversion and openness respectively (refer to Section 3.2 for details on our personality measurement). We observe a range of between 12% (contributor 12972) and 66% (contributor 2419) for Top Members' use of positive and social words, and between 15% (contributor 12972) and 35% (contributor 13664) for complex word usage respectively. On average these measures indeed correspond with our overall personality metrics, as noted in Table 3, in that the contributor 12972 measures for extroversion and openness are among the lowest for the Top Members, while contributors 2419 and 13664 are among those with the highest profiles for extroversion and openness respectively. We discuss these findings next.

*Discussion of RQ2. What are the personality profiles of Top Members?* Although we cannot conclusively link personality profiles to practitioners' engagement within teams from the results obtained here, we did find evidence of higher levels of specific personality profiles among those that engaged the most. Top Members exhibited most openness to experience, extroversion and agreeableness. The openness to experience profile is associated with being insightful and open to new ideas, those who are extroverted tend to look for stimulation from the external world, and agreeable individuals are sensitive to others [36, 40]. Such orientations are thus fitting of Top Members, and particularly given our findings in the previous section. In the position of central knowledge diffusers, a disposition that is insightful and open to others' viewpoints would encourage participation by many members [95, 96]. Similarly, a social and more sensitive outlook would be encouraging (and potentially tolerant) of others who are less capable, or those new to the team and who are needing guidance and contextual knowledge.

**Table 4. Results comparing differences in selected language usage for Top Members who contributed to multiple teams**

| Contributor | Team ID | *t*-Test: Two Sample Assuming Unequal Variance (*p*-value) | | |
|---|---|---|---|---|
| | | First-person pronouns | Social process words | Discrepancy words |
| 4661 | P1, P2 | 0.88 | 0.92 | 0.89 |
| 2419 | P1, P2 | 0.90 | 0.74 | 0.69 |
| 13722 | P3, P10 | 0.95 | 0.25 | 0.09 |
| 4674 | P3, P5 | 0.99 | 0.81 | 0.24 |
| 13664 | P6, P7 | 0.91 | 0.35 | 0.60 |

Thus, Top Members' pronounced expressions of such profiles is desirable, and particularly in globally distributed environments where distance is likely to limit the formation of team trust [97-99]. That said, and as noted above, we cannot be entirely sure that it is these personality profiles that are the drivers of Top Members' performance [16], particularly given the potential influences of expertise and task complexity on an individual's engagement [41]. The relevance of personality orientation on an individual's behavior has been established [14, 16], however, and so the evidence that most active developers behaved in socially desirable ways is relevant.

In fact, we observed the openness to experience personality profile being most pronounced for Top Members. This finding is somewhat convergent with those of other work [51]; however, these authors examined the StackOverflow forum, as against software development teams. Given that software developers frequently share their insights through the StackOverflow medium, this convergence in evidence may be indicative of a specific psychological orientation of top software developers [14]. Could it be that highly knowledgeable software developers are generally most open to experience? Perhaps the possession of such an expansive knowledge base encourages openness and tolerance? Earlier evidence from OSS examinations tends to contradict these propositions. In fact, the opposite outcome was reported in previous work conducted using the LIWC tool on the Apache mailing list [50]. Caution should be taken in assessing this divergence, however, as earlier work [87] had found the Apache group to be very selective in accepting contributors, and this may have affected the findings observed by these authors. Our divergent findings may also indicate a difference in teams' membership working in varied distributed software development environments, or differences related to global development domain characteristics and strategies. Of course, although only considering blog posts, outcomes of recent work also converges somewhat with our findings [51], lending some confidence to our proposition.

For instance, in OSS settings members join and leave software projects due to their personal motivation and, in general, informal mechanisms are used for selecting contributors [100]. On the other hand, practitioners in commercial projects are tangibly rewarded for their efforts and are selected using more formal processes. Similar to the likely adoption of policies to promote knowledge redundancy, successful commercial software organizations (such as IBM Rational) are likely to implement human resource strategies that would ensure intense screening of selected practitioners (although this is not to say that all OSS developers are not properly screened before being allowed to contribute to a codebase). In such organizations, most software-related positions demand multiple capabilities, including intra-personal, organizational, inter-personal and management skills [101, 102].

Thus, the differences observed in contributors' behaviors across OSS and commercial settings may also be deep-seated in the way practitioners for these projects are selected and assigned. Additionally, formal strategies for team building and maintaining team harmony (such as workshops and training courses) are implemented in commercial organizations, and these may also have impacted the different results observed in this study. Our findings in this case provide some support for such propositions, and particularly when considered in relation to our results for RQ1. We found that, apart from consistency for higher openness to experience profiles, highly active members (as a group) expressed all the Big Five personalities; while some Top Members were extroverted, others were agreeable, and others expressed neuroticism, with the conscientiousness personality profile being least evident. This range of profiles could be valuable in helping to balance team climate.

Our results above for the diversity of social, positive and complex word use by Top Members also support the diversity noted in personality profiles. Furthermore, when this evidence is examined in relation to Top Members' work and achievement profiles (considered in Section 4.1), the argument about these members' personality perhaps driving their performance becomes somewhat plausible. Thus, to provide further insights into this phenomenon we examine the personality profiles of those who are less active (Others) and compare their profiles with those of Top Members in the next section.

## 4.3 Top Members' versus Others' Personality Profiles

We examine the personality profiles of those who are less active and compare these measures to those of the Top Members. This assessment allows us to further assess the likely role of personality on Top Members' performance in knowledge diffusion. Table 5 shows that all of the personality profiles were evident in the discourses of those participating moderately across the ten teams. However, measures for extroversion were the highest among these members (see column three, Table 5). Openness to experience and agreeableness were also pronounced among these individuals (columns four and five), with neuroticism being less evident, and particularly when compared to the Top Members. Our ANOVA test result indeed confirmed that the range of personality profiles among these members was significantly different ($p < 0.01$), with extroversion tending to dominate.

Overall, our *t*-tests confirmed that while those contributing fewer messages and task changes also displayed the openness to experience personality profile (e.g., see Open measures for P1, P3 and P5 in Table 5), much higher levels of this profile were evident among the Top Members, and this difference was statistically significant ($p < 0.01$). Generally, Top Members also exhibited greater levels of neuroticism, extroversion, and agreeableness than the less active members, but these differences were not statistically significant ($p > 0.05$). On average, the less active practitioners (Others) were slightly more conscientious, but again, we did not find a statistically significant difference when the two groups were compared ($p > 0.05$).

Similar to the process used in the previous section, we examined less active members' usage pattern of positive and social process words (e.g., "give", "buddy", "love", "explain", "friend", "beautiful", "relax", "perfect", "glad" and "proud") and complex words (e.g., words with more than six characters in length or those with multiple syllables) to triangulate our findings. We observe a range of between 15.6% (for P6) and 41.6% (for P10) for Others' use of social and positive words, and between 18.9% (for P6) and 23.8% (for P3) for complex word usage respectively. These values, although lower than those of Top Members, are more consistent with those for the more active contributors.

We next examined the differences in the individual personality profiles for the Top Members and Others involved in the teams (P1–P10) to see how personality profiles were distributed within these teams and for those involved in specific tasks. A subset of these results (those that are most interesting) is represented in Figure 5(a-b). Overall, in each graph (a-b) we observe that for all the personality profiles, when Top Members' scores were high, scores of the less active contributors were low. In addition, this pattern was reversed when the less active contributors had more pronounced personality profiles. For example, in Figure 5(a) Top Members working in teams P1, P2, P3, P6 and P9 demonstrated higher levels of extroversion than the less active members, while the opposite was seen for this personality profile in the remaining teams (P4, P5, P7, P8 and P10). Similarly, Figure 5(b) shows that while Top Members typically express more pronounced openness to experience (as this pattern was evident for six of the ten teams, with two teams being relatively even), the opposite pattern was seen for the less active members of teams P1 and P7. This inverse personality pattern was also maintained for the other profiles.

Finally, we removed the values for the practitioners that were not initially selected in our Top Members cluster given their less than 0.33 density measure, and performed further comparisons to validate our findings. These members include practitioners 4674, 12702 and 2102 (refer to Table 2 for details). Note in Table 2 also that a number of members were active on multiple projects, and so, of the 14 members that were initially selected to the Top Members cluster given their satisfaction of the 0.33 density measure threshold, 12 were distinct. We performed formal pair-wise comparisons (*t*-tests) to see if our pattern of results above would hold given these omissions (i.e., the removal of measures for contributors 4674, 12702 and 2102). Our *t*-tests confirmed that of the five personality profiles, only openness to experience recorded a statistically significant difference ($p < 0.05$) in favor of Top Members, replicating the results presented above. This finding confirms that openness to experience was the only differentiator of Top Members' and Others' personality profiles. We discuss these findings in relation to theory next.

*Discussion of RQ3. Do the personality profiles of Top Members differ to those of practitioners who are less active?* Apart from the openness to experience personality profile, the personality profiles of Top Members were not significantly different to those of their less active (Others) counterparts. However, we are not able to conclusively link the openness to experience personality profile to the dominant presence of Top Members, and practitioners' involvement in knowledge diffusion more generally, given our single snapshot study. That said, capacity (e.g., skills and intelligence) and opportunity (e.g., working conditions)

notwithstanding, personality and other psychological factors (e.g., motivation and self-image) have been shown to impact both the judgment and willingness of individuals to perform [14, 103, 104]. Such variables may help us to predict the future actions of those being assessed. In fact, the openness to experience trait under consideration here is often associated with intelligence [14], and so, Top Members' higher scores on this dimension is perhaps not very surprising. Others have indeed confirmed that intuitively inclined individuals tend to outperform those who are less so [105].

**Table 5. Other contributors' personality scores (values are relative not absolute)**

| Team ID | Neuro | Extro | Open | Agree | Consc |
|---|---|---|---|---|---|
| P1 | 6.40 | 47.31 | 49.77 | 40.22 | 1.26 |
| P2 | -6.41 | 52.50 | 33.93 | 27.18 | 3.62 |
| P3 | 9.02 | 41.19 | 40.73 | 31.97 | 5.91 |
| P4 | -8.45 | 51.42 | 36.30 | 28.60 | 4.14 |
| P5 | 18.72 | 38.18 | 46.31 | 36.46 | 6.89 |
| P6 | 30.87 | 20.79 | 32.81 | 36.58 | 9.94 |
| P7 | 11.03 | 38.67 | 37.82 | 35.06 | 3.69 |
| P8 | -5.30 | 55.34 | 36.59 | 26.75 | 3.63 |
| P9 | 28.82 | 23.17 | 32.68 | 35.73 | 9.38 |
| P10 | -7.24 | 50.15 | 40.38 | 28.94 | 5.40 |

We observed that as a collective group the Jazz developers were very extroverted, agreeable and open to experience, and overall, all teams were heterogeneous (i.e., members collectively exhibited all of the Big Five personality profiles). Interestingly, for all ten teams investigated, when Top Members scored high on some profiles those who were less active scored lower on those profiles, while the opposite was seen when less active members scored higher for specific personality profiles. These patterns seem to indicate that there may be some form of self-organization of behaviors among the leaders and those less active in Jazz teams [75, 106]. Of course this evidence may also be linked to a deliberate human resource strategy that is employed by IBM Rational. In fact, in considering the linguistic dimensions that are used for constructing the personality metrics, it would be fitting for such variations in attitudes to be expressed among Jazz practitioners. Our findings point to the relevance of understanding teams' behaviors in order to fully comprehend the properties of their outcomes.

For instance, a continuous stream of insightful words (e.g., think, know, consider) that also characterize those who exhibit openness to experience may be useful for maintaining team ideas, and so it would be fitting for those who are less active to express such terms when there is a reduction of such discourses from Top Members. At this time specific (Other) individuals may take on temporary champion roles within the teams. Similarly, a continuous stream of the positive (e.g., beautiful, relax, perfect) and social (e.g., give, buddy, love) words that characterize the content that is communicated by those that are extroverted would be beneficial for maintaining team optimism, perhaps at times of schedule pressure when members are low on morale. Thus, alternating the use of such expressions (whether deliberate or otherwise) would no doubt help with sustaining team satisfaction and morale.

In fact, although not universally accepted, previous theories have encouraged personality heterogeneity in teams, in order for teams to benefit from such a balance of behaviors. Trimmer et al. [54], for instance, noted that personality diversity boosts team confidence and satisfaction. Thus, there is theoretical support for Jazz team members' varied personality profiles, which may explain why the tools developed by these teams have recorded such high usage and positive reviews (refer to jazz.net). That said, the evidence noted for the variation of attitudes through the study of multiple teams has not been provided previously.

We correlated the personality profiles and in-degree, density and closeness scores for Top Members and Others to see if there were statistically significant relationships between these dimensions (e.g., to ascertain if those practitioners who demonstrated more openness to experience and extroversion profiles were more influential in knowledge diffusion). The relationship between apprehension (or lack thereof) about participating in team communication has long been shown to be linked to the personality profiles of individuals [53], and so establishing such a linkage for software practitioners could be useful for informing team composition strategies. We performed Pearson product-moment correlation tests, which did not return any statistically significant results ($p > 0.05$) – the best result being $r = -0.31$, $p = 0.13$ for the neuroticism profile and the in-degree measure (i.e., individuals with higher neuroticism profiles had lower in-degree scores). We are therefore not able to link personality profile to knowledge diffusion (or apprehension about participating in team communication) in this work.

We next consider IBM Rational Jazz teams' personality profiles more generally in Section 4.4. This final stage of analysis is used to validate our findings against those of others who previously examined the personality profiles of software practitioners.

## 4.4 Jazz Teams' Personality Profiles

In this final section of the results we examine Jazz teams' personality profiles, using the team as our unit of analysis. Given the relatively small sample of teams, we do not claim generalizability of our findings, but instead compare these to the findings of others to assess convergence.

Overall, measures for extroversion were the highest for the ten IBM Rational Jazz teams studied in this work (teams' composite score: mean = 43.4, median = 39.9, std dev = 13.7). However, Jazz practitioners also appeared highly open to experience (teams' composite score: mean = 41.2, median = 39.6, std dev = 6.5) and agreeable (teams' composite score: mean = 33.4, median = 36.1, std dev = 7.0). On the other hand, these practitioners were least conscientious (teams' composite score: mean = 5.3, median = 5.4, std dev = 2.4), and moderately neurotic (teams' composite score: mean = 10.0, median = 13.2, std dev = 20.8).

Given the patterns observed in Figure 5(a-b), Pearson product-moment correlation testing was undertaken to determine how the different personality profiles varied among the 146 Jazz practitioners. Notwithstanding the psycholinguistic approach used in this work as against the more commonly used questionnaire-based instruments [9, 30], we checked to see how IBM Rational Jazz teams' personality profiles compared to those of other software practitioners who had previously completed the Big Five questionnaire [107]. Our findings in Table 6 show that those Jazz practitioners who exhibited extroversion did not display the neuroticism, agreeableness and conscientiousness profiles; these results are strong negative correlations, and are statistically significant ($p < 0.01$). On the other hand, Table 6 demonstrates that the most conscientious practitioners tended to also be agreeable and demonstrated some degree of neuroticism. These are medium and strong, statistically significant correlations ($p < 0.05$ and $p < 0.01$ respectively). Similarly, those Jazz practitioners who were most agreeable also demonstrated openness to experience and neuroticism; these are strong positive correlations, and statistically significant ($p < 0.05$ and $p < 0.01$ respectively).

In terms of the way personalities varied across individual teams conducting different forms of tasks, while all Jazz teams exhibited similar profiles for openness to experience and agreeableness, those working on user experience tasks exhibited higher levels of extroversion than the teams involved with documentation, coding and project management tasks (refer to Table 1). Formal statistical testing (*t*-tests) confirmed that these differences were statistically significant ($p < 0.01$). Those working around documentation and project management tasks also exhibited more extroversion than coders. On the other hand, teams working on coding-intensive tasks were most neurotic and conscientious; these differences were also statistically significant ($p < 0.01$).

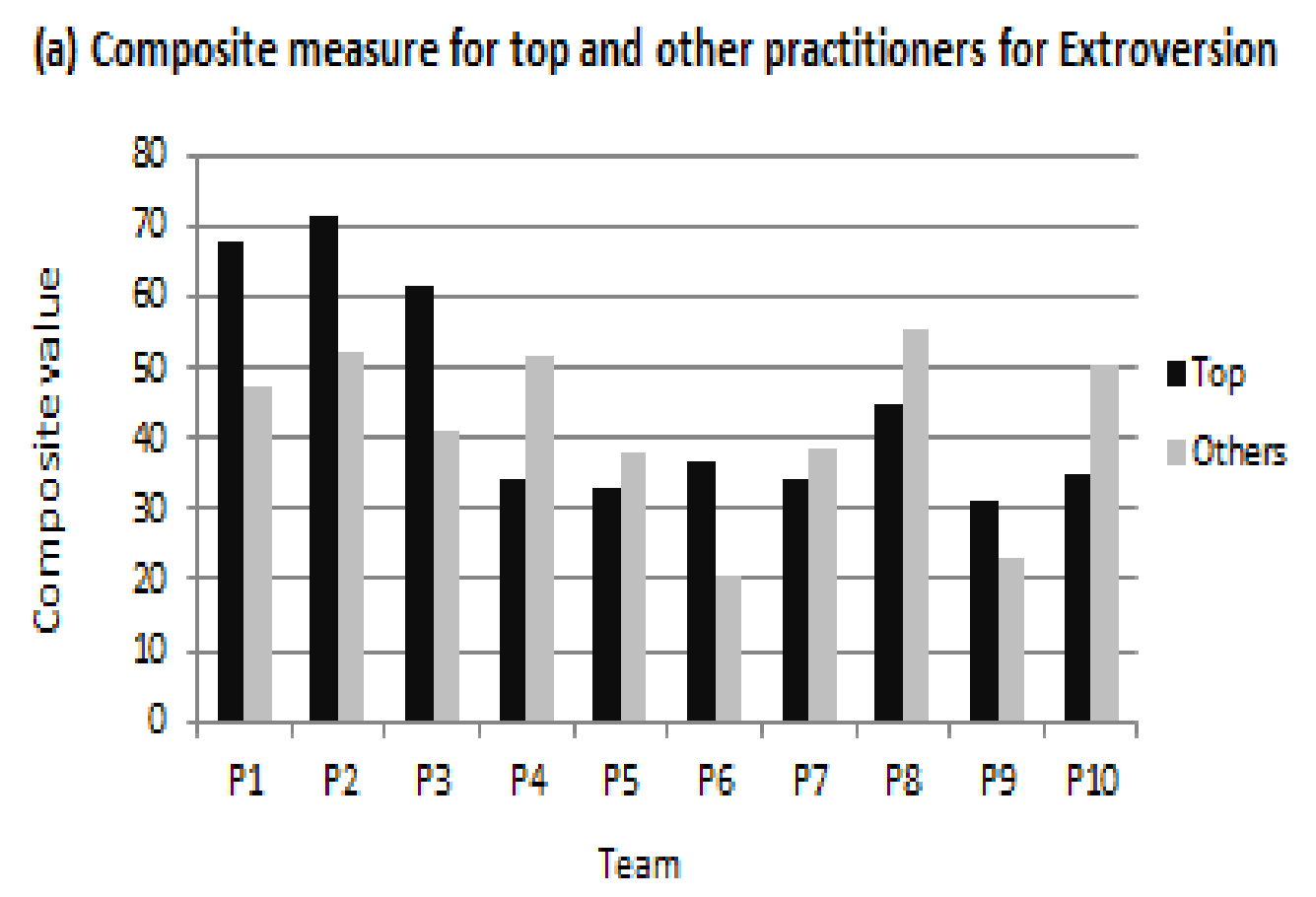

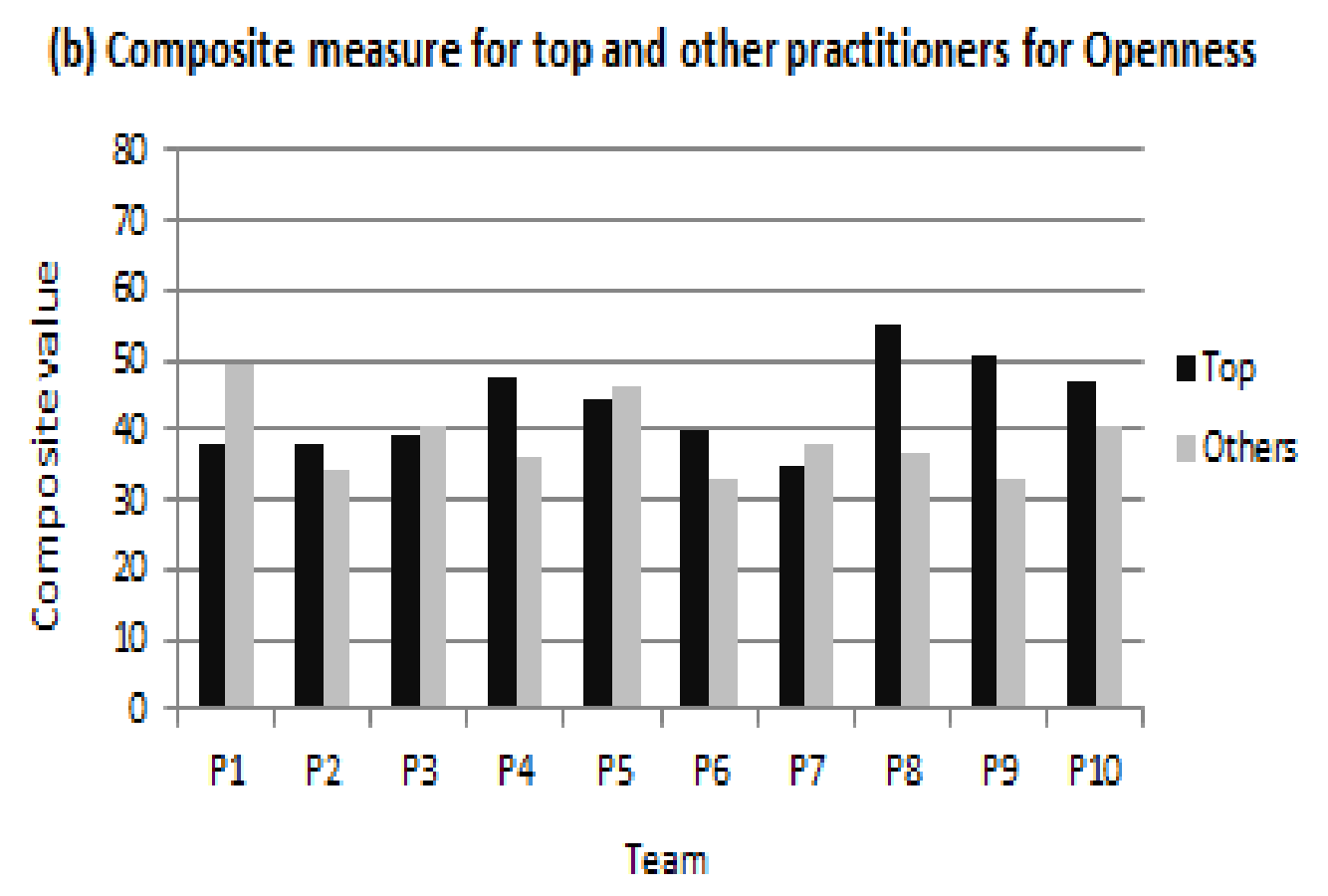

**Figure 5. Average personality score, Top Members[7] and Others (values are relative not absolute)**

**Table 6. Correlations (*r*) for the way personalities varied**

| Personality Profile | 1 | 2 | 3 | 4 | 5 |
|---|---|---|---|---|---|
| 1 Neuro | 1.0 | **-0.929**** | 0.279 | **0.915**** | **0.647**** |
| 2 Extro | | 1.0 | -0.086 | **-0.852**** | **-0.775**** |
| 3 Open | | | 1.0 | **0.511*** | 0.063 |
| 4 Agree | | | | 1.0 | 0.494* |
| 5 Consc | | | | | 1.0 |

**Note**: bold font denotes strong correlation, *p < 0.05, ** p < 0.01

*Discussion of RQ4. How are personality profiles distributed in a successful global team?* Overall, these global practitioners mostly expressed themselves with extroversion, but openness to experience and agreeableness profiles were also pronounced in the discourses of Jazz members. In fact, the varied distribution of personality profiles in this study is similar to observations made regarding the 130 software engineers who completed the Big Five instrument in the study reported by Sach et al. [107]. This evidence suggests that as a group, software developers may share some particular characteristics regardless of the environment in which they are operating – whether in distributed or collocated settings [104]. In addition, this finding also encourages confidence in our analysis and outcomes.

From a task perspective, and notwithstanding our small sample size, extroversion was most distinct amongst the group of developers undertaking user experience oriented tasks, and may signal that there is a general tendency (or preference?) for individuals in such an environment to be driven by others. In fact, we previously found such members to consistently express a large amount of positive language [45], somewhat endorsing our proposition. Given that usability practitioners are often required to conduct user experience evaluations involving the wider team, both for assessing ease of use and ensuring that features match previously planned requirements, individuals aligning to a more extroverted outlook seem fitting for these tasks. Such an outlook may also serve to be useful for lobbying and encouraging fragmented members in GSD teams.

In fact, Gorla and Lam [15] found high performing programmers to be extroverted, and extroversion was observed to be a positive project management profile [16] – particularly for effective communication. We also found pronounced extroversion among project managers, somewhat endorsing the view that those leading software teams (and particularly in distributed development contexts) may benefit if they possess excellent inter-personal skills. Such skills would promote openness and trust, critical ingredients for establishing rapid team identity and performance [81, 82] in a GSD environment where *trust without touch* is required.

We found those involved in coding tasks to be more conscientious. However, there is some uncertainty over whether these results are linked to the nature of the tasks they performed or the coding teams' actual expertise [41, 46]. Previous evidence had indeed found an individual's capability and the challenge presented by the task(s) they are performing to be stronger predictors of their diligence than their actual behavior orientation [41]. That said, the notion that teams and individuals are often affected by the tasks under their purview has also been previously supported by empirical evidence [84]. In fact, we also observed a higher incidence of neuroticism (or negative language) among those involved in coding tasks, an observation we believe could be linked to the rigor required in solving challenging computational tasks [46]. Overall, however, our results suggest some uncertainty regarding the personality profile of neuroticism, as we did not find full agreement with prior psycholinguistic theories. For instance, contrary to findings in the psycholinguistic space, we found those expressing negative emotion (e.g., afraid, hate, dislike) to also use articles (e.g., a, an, the); an outcome also encountered by others [42, 43].

[7] These measures are not to be confused with those in Table 3, whose measures are for individual Top Members overall, and not their measures on the specific projects (P1 – P10).

Given the foregoing divergence in previous outcomes and the challenges encountered when assessing the neuroticism personality dimension, we are thus not able to conclusively link coding tasks to the expression of higher levels of conscientiousness and neuroticism.

# 5. IMPLICATIONS

The outcomes of this work have several implications, for research design and theory, and in relation to global software development practice. We summarize these implications in the following two subsections. We first consider how this study's design and outcomes may provide insights for those performing similar future studies and software engineering theory in Section 5.1. Section 5.2 then outlines the implications of our findings for the governance of globally distributed software development projects.

## 5.1 Implications for Research

From a methodological or design perspective we have employed multiple techniques in this work, for partitioning the data, selecting our study participants, and for analyzing the study data. The incremental way in which the results were uncovered using the different techniques (data mining, SNA, and linguistic analysis) shows that these procedures, if applied systematically, may complement each other and strengthen the validity of research concerned with behavioral issues. This need to be systematic is particularly necessary when conducting research based on secondary data drawn from repositories, as opposed to being collected by the researcher in a live setting.

In terms of implications for theory, from the evidence uncovered in this work and that also emerged in our separate study [52], we posited that Top Members in this study were most important to knowledge diffusion. Given the single context considered, however, research should look to replicate this study and explore such members' communications using other deeper approaches, perhaps involving thematic analysis, to provide confirmation for our findings. Additionally, the SNA conducted in this research shows that teams with centralized and low-density communication networks may also remain highly connected through their collaboration on software tasks. We believe that such connected communication networks may aid with knowledge diffusion. However, there still remain questions around the effectiveness of such a strategy for dealing with knowledge transfer. Accordingly, future research is encouraged to study this issue.

Furthermore, while previous works have linked various personality profiles to team behavior and performance through the use of questionnaire-based techniques and for collocated developers [14, 16], here we studied the personality profiles of distributed global teams using practitioners' textual communication, and have uncovered slightly divergent findings to those uncovered for OSS teams [50]. This suggests that there may not be universally successful personality configurations as such, given that, although our findings are somewhat divergent to those that were uncovered for Apache developers, the software products deployed by both teams may be considered to be highly successful.

That said, our findings may also be moderated by specific work practices employed by the teams studied, as we noted other patterns that were not typically observed by others. Furthermore, issues associated with data capture in OSS projects [78] may also have contributed to the divergence in findings. We thus encourage the execution of similar studies of other globally distributed teams to validate our outcomes.

## 5.2 Implications for Practice

Our study also has implications for managers. In this study Top Members were the only communicators around a relatively large number of software tasks; a pattern that we believe represents a risk if evident for similar distributed teams. In fact, the threat imposed by the loss of key team players (and with them – the team's tacit knowledge) has been a recognized source of concern for agile teams, in general, given their reliance on team members' interaction as a substitute for extensive documentation [108-110]. This threat is likely to be exacerbated for globally distributed software teams, where there are limited possibilities for spontaneous and informal communications [111], and the need to document events and practitioners' action may be a burden. Further, such environments also present social challenges for members involved [20, 112]. Accordingly, in making provision for Top Members' (or any other practitioners') absence or sudden withdrawal from the team, project management in a global agile setting may promote team configurations (and collaborations) that are likely to provide failsafe mechanisms (e.g., task assignment that promotes some level of knowledge redundancy). Such team configurations are likely to reduce the threat imposed by the loss of key team players, and with them, the team's tacit knowledge.

We confirmed that Jazz's main contributors were most open to experience, but that all personality profiles were evident during team work. Those who express the openness to experience personality profile are said to be insightful and are often receptive to new ideas [40]. Individuals who naturally possess these behaviors may be integral to their teams' performance, and particularly if they occupy the center of their teams' communication networks. Team extroversion and agreeableness were also seen to be evident among Jazz members. Individuals with the agreeableness profile have been shown to be cooperative, compassionate and sensitive to others. Additionally, extroverts are social and are driven by the external world. These behaviors may collectively be useful for maintaining teams' synergies, particularly in global settings where there are limited avenues for face-to-face communication and rapid development of team trust.

Although previous work has posited that personality heterogeneity has little impact on team performance [15], variations of personalities may have a balancing effect during distributed team work. We observe this personality diversity (and variances in practitioners' discourses) among Jazz distributed teams, although there is still some uncertainty about whether our observations are linked to a deliberate IBM human resource strategy, or if these patterns naturally emerge because these are self-organizing teams of highly skilled individuals. For instance, negative emotion has been shown to affect team cohesiveness and is linked to individualistic behaviors and neuroticism [16], while use of positive and social language has the opposite effect and is associated with extroversion [40]. Thus, in times of work intensity and during stressful situations when negative feelings are festering, the more social and agreeable attitudes may help to mitigate conflict and to maintain team optimism. Accordingly, project managers may encourage team members exhibiting these more positive attitudes at times when their teams' morale is low or when there is need for persuading and encouraging involvement from the members of the wider software organization. Similarly, the more aggressive and conscientious behaviors may be useful for promoting team urgency, and may generally be allowed during periods of frustration when team deadlines are approaching or during complex computational work.

Software project managers may use these trends to assemble and manage teams with individuals who are together both socially and mentally equipped for any given software situation.

# 6. THREATS TO VALIDITY

*Construct Validity:* The language constructs used to assess personality in this study were used previously to investigate this phenomenon and were assessed for validity and reliability (e.g., [39, 50]). However, the adequacy of these constructs, and suitability of the LIWC tool for studying software practitioners' linguistic processes, may still be subject to debate. Communication was measured from messages sent around software tasks. Although project communication is encouraged through the use of Jazz [113], these messages may not represent all of the teams' communication. Additionally, cultural differences may have an impact on individuals' behaviors; however, research examining this issue in global software teams has found few cultural gaps among software practitioners from, and operating in, Western cultures, with the largest negative effects observed between Asian and Western cultures [112]. Given that the teams studied in this work all operated in Western cultures, this issue may have had little effect on the patterns of results observed. Furthermore, we posit that Jazz teams are successful based on usage of the RTC (see jazz.net); however, this measure does not account for within-project success indicators (e.g., those related to schedule and budget).

*Internal Validity:* Although we achieved data saturation (refer to Section 3.1) after analyzing the third set of team networks (and all teams were homogenous – refer to Appendix A for the Jazz teams' sociograms showing centralized communication networks), the history logs and messages from the ten teams may not necessarily represent all the teams' processes and activities in the repository.

*External Validity:* The work processes at IBM are specific to that organization and may not represent the organization dynamics in other software development establishments, and particularly for environments that employ conventional waterfall processes [114]. Such environments may employ more rigid project management practices, with clear hierarchical structures and development boundaries [115, 116]. Accordingly, in such conventional teams Top Members are not likely to exhibit the dominant presence noted across such a large volume of software tasks. In fact, Jazz organization culture may also have impacted the communication patterns noted, including the way some programmers occupied central roles. That said, Costa et al. [117] confirmed that practitioners in the Jazz project exhibited similar coordination needs to practitioners of four projects operating in two distinct companies. Thus, we believe that our results may be applicable to similar large-scale distributed projects.

# 7. CONCLUDING REMARKS

It has long been posited that software repositories and interaction logs possess evidence of team relations that may be revealed through systematic observations of these artifacts. One such revelation is that few members dominated team communication during software development. However, little is known about these members role in project knowledge diffusion. Similarly, while a large volume of research has been dedicated to understanding software developers' personality traits, there has been little focus on studying the potential influence of personality in distributed and GSD teams. In particular, researchers have placed scant attention to the examination of personality profiles of top members, and evaluating how such profiles affect the willingness of practitioners to be involved in knowledge diffusion.

Accordingly, in this study we employed data mining, SNA and linguistic analysis to explore software practitioners' communication and personality using artifacts from IBM Rational Jazz global development teams. Our findings confirmed that teams' Top Members were central to their teams' knowledge diffusion, these members exhibited more openness to experience, and all personality profiles were represented during team work. Additionally, we found evidence of high levels of openness to experience, agreeableness and extroversion among all teams, and the highest levels of extroversion among those working on user experience tasks. In contrast, those involved in coding tasks were the most neurotic and conscientious. These findings somewhat endorse those of previous researchers, and confirm those in the psycholinguistic space. However, our results are divergent to those that were previously established using Apache OSS mailing lists.

We suggest that these differences may be linked to the processes involved in selecting software practitioners in commercial organizations such as IBM, and the training strategies implemented to maintain standards in such organizations. That said, it would be useful to examine whether our results hold for other global teams and for those undertaking other forms of software tasks. Additionally, future studies are encouraged to triangulate our results with bottom-up analysis techniques.

# 8. ACKNOWLEDGMENTS

We thank IBM for granting us access to the Jazz repository. Thanks also to the EASE'14 reviewers for their detailed and insightful comments on the shorter version of this work that considered the treatment of team personality.

**APPENDIX A. Sociograms for selected Jazz teams (P1 – P10)**

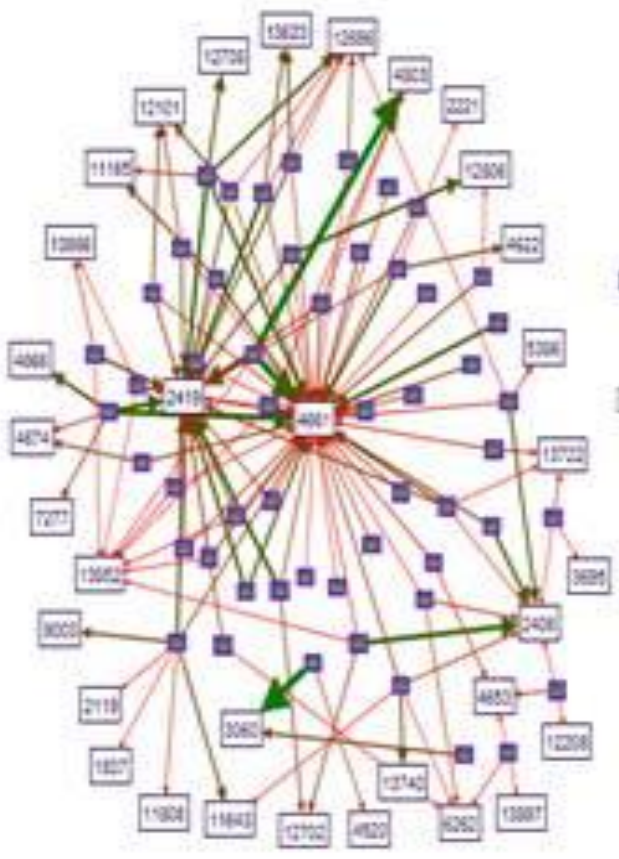

P1

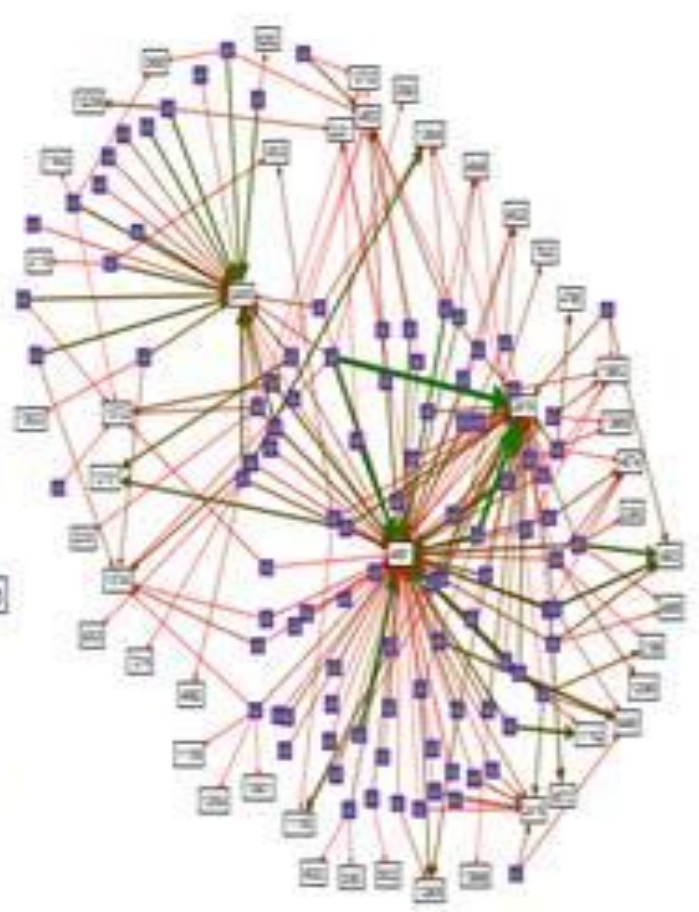

P2

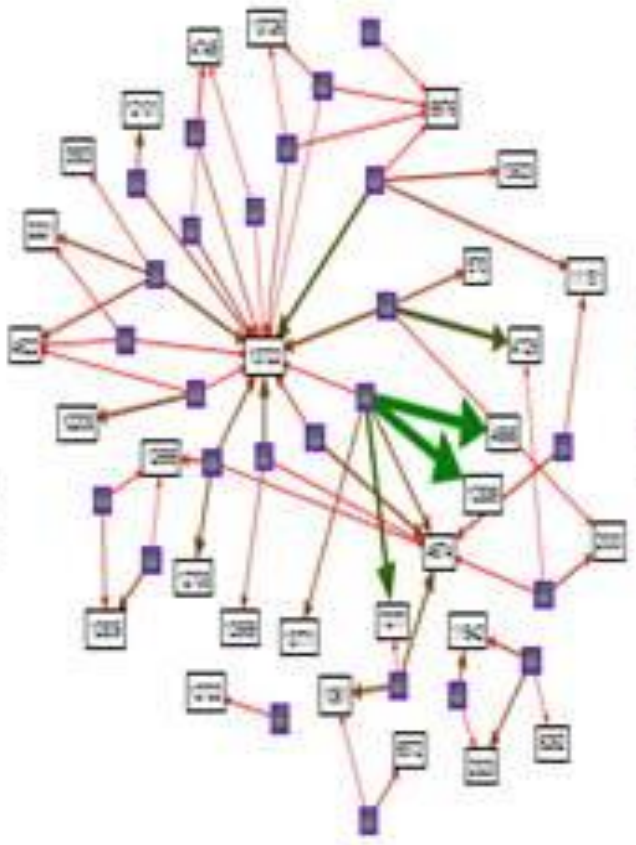

P3

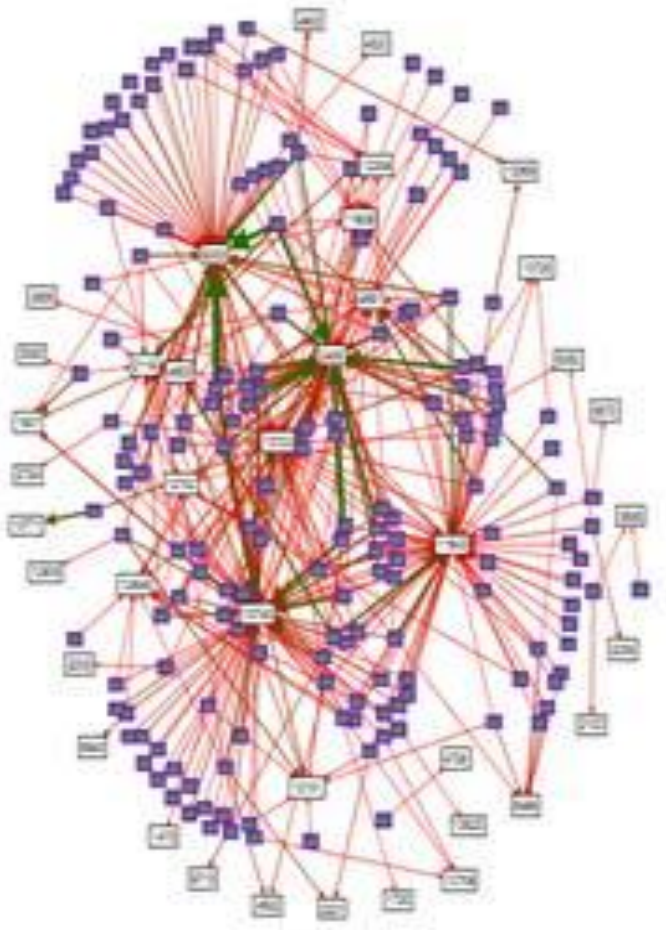

P4

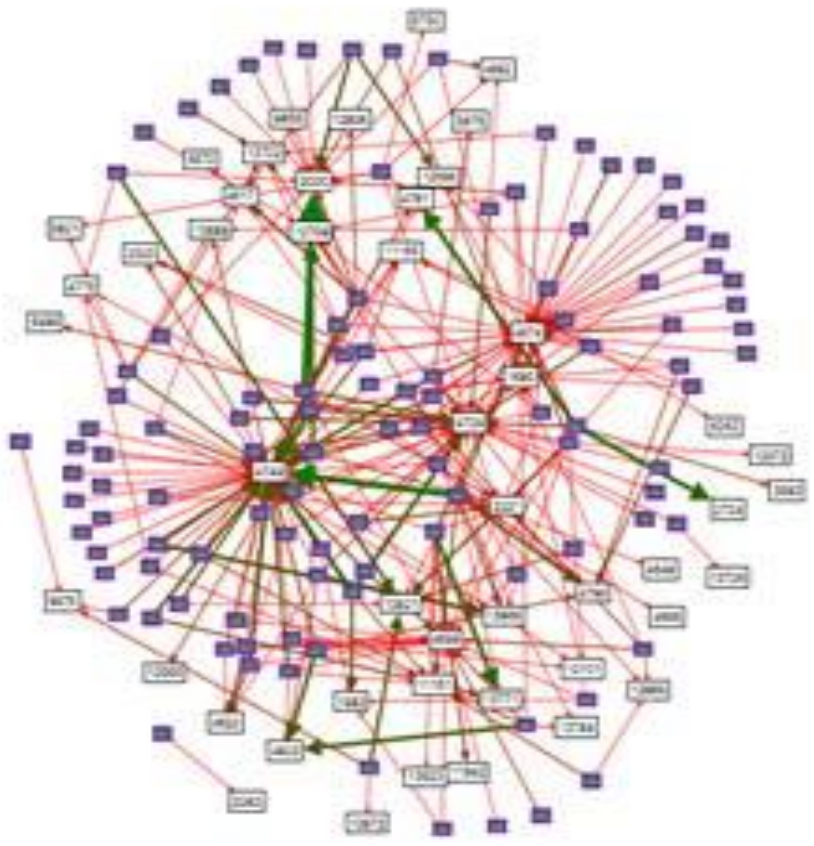

P5

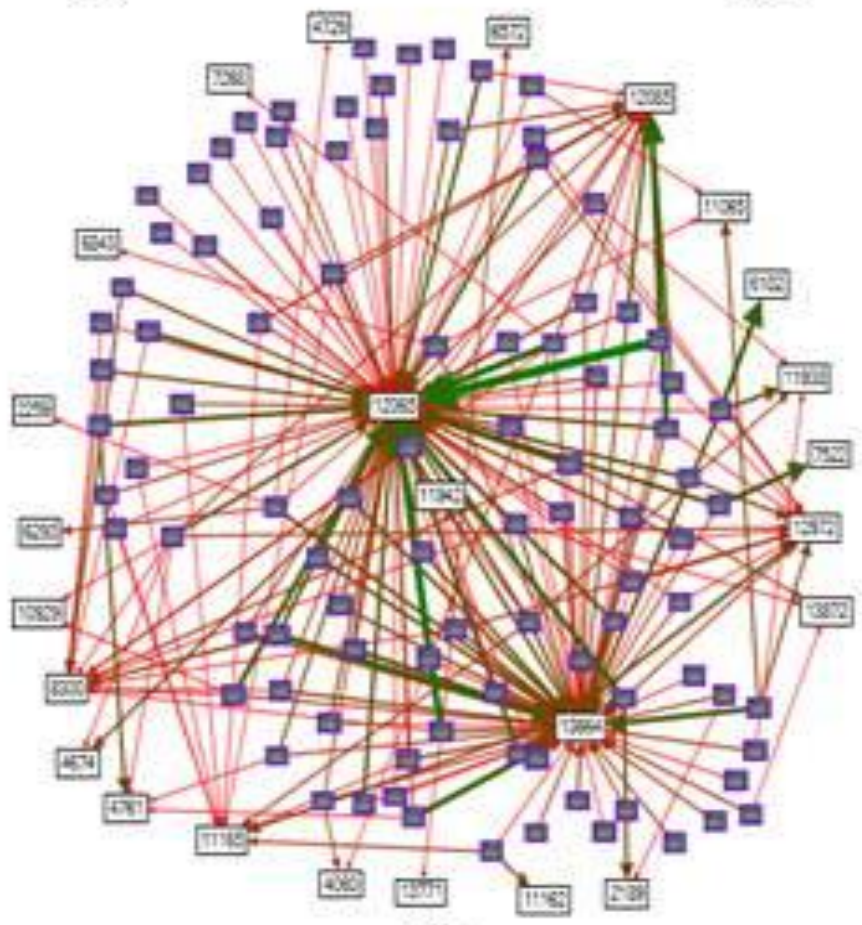

P6

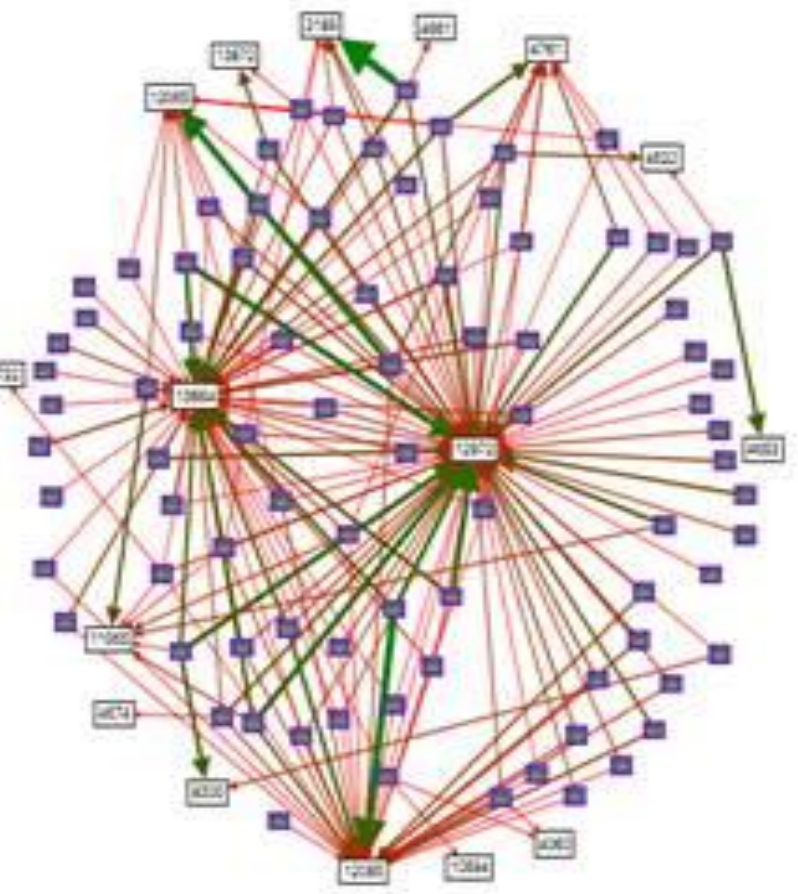

P7

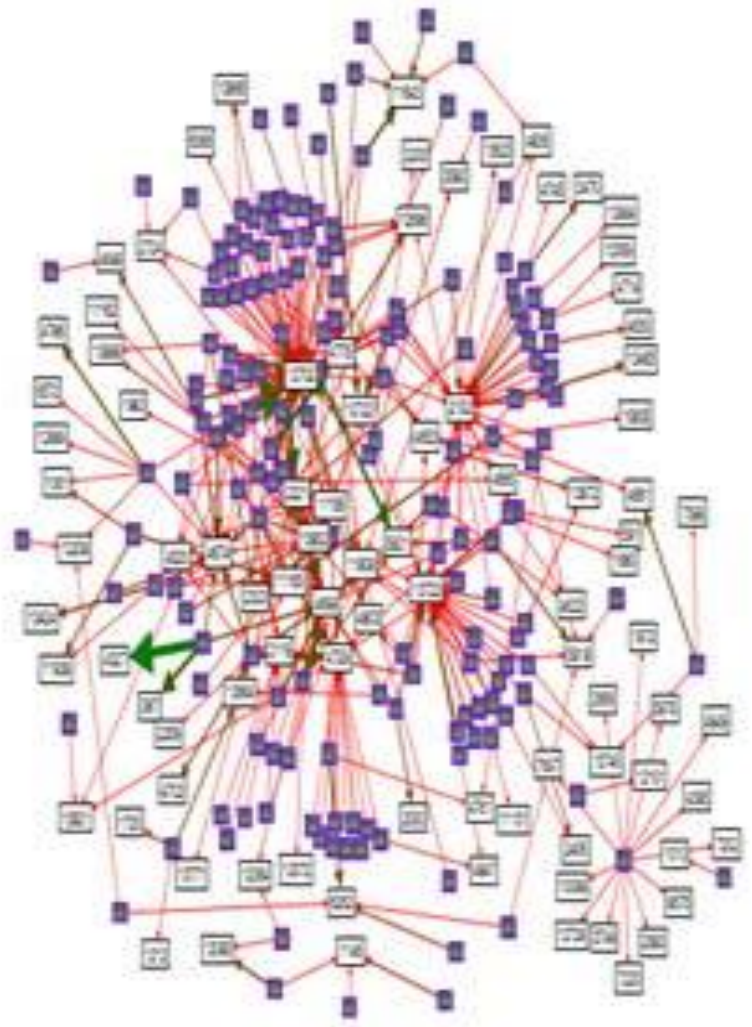

P8

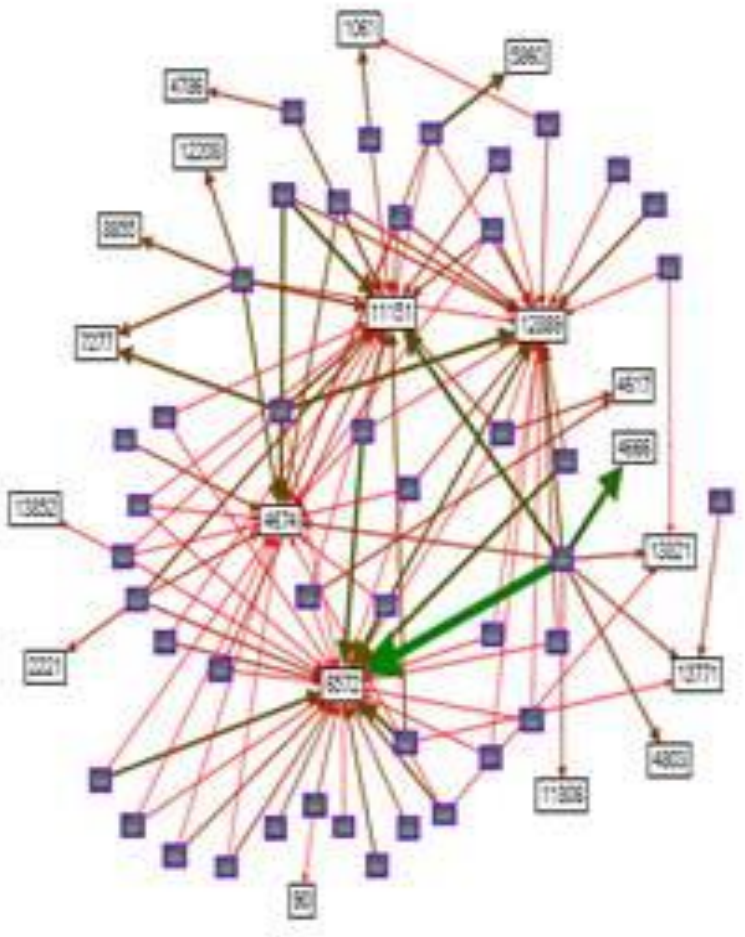

P9

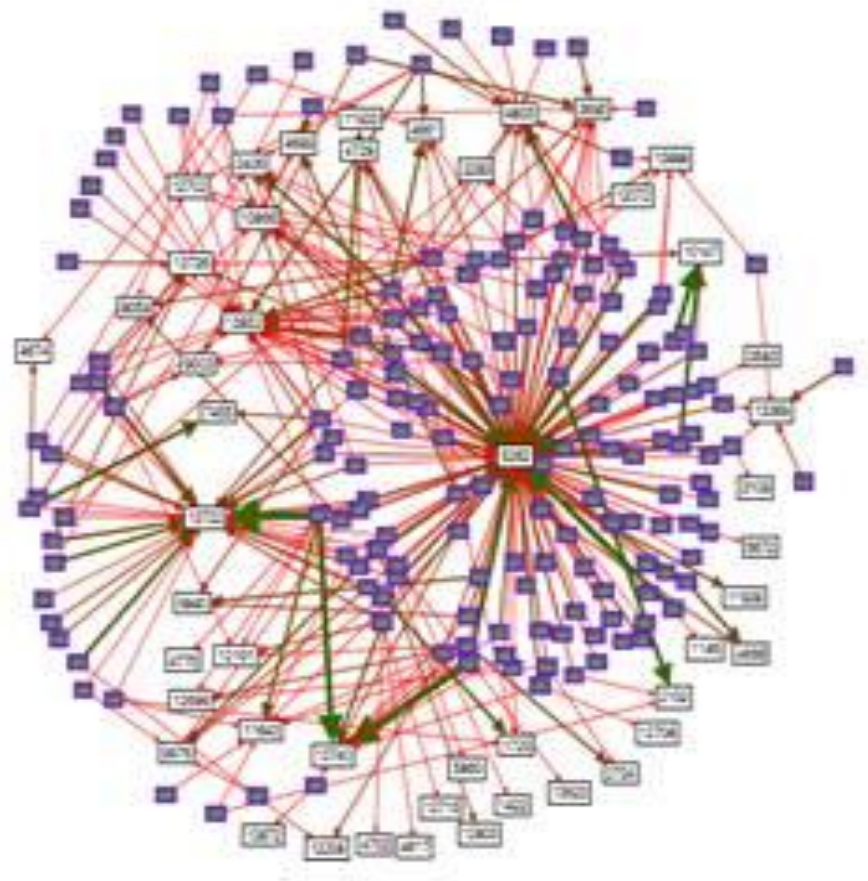

P10